\documentclass[twocolumn,twocolappendix]{aastex701} 
\usepackage{amsmath}
\usepackage{soul}

\begin{document}

\shorttitle{Dynamo-Generated Poloidal Magnetic Fields in 3D Collapsar Simulations}
\shortauthors{Chan, Gottlieb, Jacquemin-Ide, Cantiello \& Renzo}
\title{Jets from Scratch: Dynamo-Generated Poloidal Magnetic Fields in 3D Collapsar Simulations}

\author[0000-0003-2776-082X]{Ho-Sang Chan}
\altaffiliation{Croucher Scholar}
\altaffiliation{CCA Predoc Fellow}
\affiliation{JILA, University of Colorado and National Institute of Standards and Technology, 440 UCB, Boulder, CO 80309-0440, USA}
\affiliation{Department of Astrophysical and Planetary Sciences, University of Colorado, 391 UCB, Boulder, CO 80309, USA}
\affiliation{Center for Computational Astrophysics, Flatiron Institute, New York, NY, USA}
\email{hschanastrophy1997@gmail.com}

\author[0000-0003-3115-2456]{Ore Gottlieb}
\affil{Department of Physics and Kavli Institute for Astrophysics and Space Research, Massachusetts Institute of Technology, Cambridge, MA 02139, USA}
\affiliation{Center for Computational Astrophysics, Flatiron Institute, New York, NY, USA}
\email{ogot@mit.edu}

\author[0000-0003-2982-0005]{Jonatan Jacquemin-Ide}
\affiliation{JILA, University of Colorado and National Institute of Standards and Technology, 440 UCB, Boulder, CO 80309-0440, USA}
\email{}

\author[0000-0002-8171-8596]{Matteo Cantiello}
\affiliation{Center for Computational Astrophysics, Flatiron Institute, New York, NY, USA}
\affiliation{Department of Astrophysical Sciences, Princeton University, Princeton, NJ 08544, USA}
\email{}

\author[0000-0002-6718-9472]{Mathieu Renzo}
\affiliation{Steward Observatory, Department of Astronomy, University of Arizona, 933 N. Cherry Ave., Tucson, AZ 85721, USA}
\email{}


\begin{abstract}

The origin of the large-scale poloidal magnetic field required to power relativistic jets in collapsars remains uncertain. While such a field may be inherited during PNS collapse, the efficiency of this process is unclear, motivating an in situ mechanism to generate poloidal fields out of the predominantly toroidal fields produced by stellar differential rotation. We present the first 3D general-relativistic magnetohydrodynamic collapsar simulations initialized with toroidal magnetic field profiles that closely follows those of pre-collapse stellar models. As the toroidal field in the disk becomes dynamically important, it seeds the dynamo, producing coherent poloidal magnetic loops that appear at $\sim \mathcal{O}(100)$ gravitational radii and are then advected inward along paths that may deviate from the disk midplane. The resulting poloidal fields thread the black hole (BH) and launch highly variable, wobbling relativistic jets on timescales of order seconds, with the onset depending on the initial magnetic field and the plasma circularization radius. Although the jets are highly variable and misaligned with the BH spin axis, they sustain $\gtrsim 10^{50}\,\mathrm{erg\, s^{-1}}$, comparable to that inferred for long gamma‑ray bursts (LGRB). We identify magnetic-flux inversions driven by the stochastic dynamo, leading to the formation of striped jets that could be imprinted in LGRB light curves. These results demonstrate that accretion-disk dynamos provide a robust pathway for jet production in collapsars across a broad range of progenitors.

\end{abstract}




\section{Introduction} \label{sec:intro}

The collapse of rapidly rotating massive stars (collapsars) is a widely accepted model for the production of long gamma-ray bursts \citep[LGRBs;][]{1993ApJ...405..273W,1999ApJ...524..262M} --- the most luminous electromagnetic transients in the Universe. In collapsars, the iron core of a massive star first undergoes gravitational collapse to form a proto-neutron star (PNS). If the PNS subsequently accretes sufficient mass from the infalling stellar gas, it collapses further to form a black hole (BH). The BH may launch highly collimated, ultra-relativistic jets \citep[e.g.,][]{1992ApJ...395L..83N,1999ApJ...525..737R} via the Blandford–Znajek (BZ) mechanism \citep{1977MNRAS.179..433B}. Once these jets break out from the star, they generate the observed LGRB emission.

The BZ mechanism requires a coherent, large-scale, dynamically important poloidal magnetic field threading a spinning BH. In the presence of such fields, magnetic field lines tap into the rotational energy of the BH and extract it in the form of outward-directed Poynting flux \citep{2003PASJ...55L..69N,2006MNRAS.368.1561M,2011MNRAS.418L..79T}. Despite its central importance, the origin of the large-scale poloidal magnetic field threading the BH remains largely unknown.

Early studies often assumed that the large-scale poloidal magnetic field originates from the progenitor star \citep[e.g.,][]{2007ApJ...664..416B,2009MNRAS.397.1153K,Mosta2014,2020MNRAS.492.4613O,2022MNRAS.510.4962G,2022ApJ...933L...9G,2023A&A...677A..19J,2024PhRvD.109d3051S,2025ApJ...985L..26I,Urrutia2025}, which is amplified during stellar collapse via flux-freezing \citep{1970ApJ...161..541L}. However, stellar evolution models show that the progenitor star primarily generates toroidal magnetic fields through the Tayler-Spruit dynamo \citep[TSD, ][]{Tayler1973,2002A&A...381..923S,Fuller2019}. The associated poloidal component is small-scale and randomly oriented, leading to a substantial cancellation of the field upon collapse. Consequently, the net poloidal flux available to thread the newly formed BH is orders of magnitude lower than that required to power an LGRB jet \citep{Gottlieb_2024}. This disparity suggests that another mechanism must be at work to generate a large-scale poloidal magnetic field.

\citet{Gottlieb_2024} proposed that the BH acquires the poloidal magnetic flux required to launch an LGRB jet from the PNS --- prior to the PNS collapse, dynamo processes in the PNS amplify the magnetic field to proto-magnetar strengths \citep{2020SciA....6.2732R,2023MNRAS.526L..88B,2023PhRvD.108l3012M,2025A&A...695A.183B,2024NatAs...8..298K,2025NatAs...9..541I}, generating the large-scale poloidal field that will thread the newly formed BH. During the PNS phase, the high-angular-momentum infalling gas forms a centrifugally supported disk. Upon the PNS collapse, the plasma currents in this disk anchor the magnetic field lines and prevent the field from rapidly decaying. Whether the BH accumulates magnetic flux faster than it is lost through electromagnetic decay depends on whether the disk remains magnetically connected during the PNS collapse; this requires the disk to viscously extend inward through the magnetospheric radius set by the strong PNS field before the magnetic flux is lost through reconnection at the horizon \citep{2021PhRvL.127e5101B}.

The model by \citet{Gottlieb_2024}, however, is subject to several uncertainties. First, the amplification timescale of the PNS dynamo may be comparable to the PNS collapse timescale, raising the question of whether sufficient poloidal magnetic flux can be generated prior to collapse. Second, it remains unclear how much of the amplified field survives the transition from a PNS to a BH, as magnetic reconnection and dynamical reconfiguration during the PNS collapse may reduce the net flux that would be inherited by the BH \citep[e.g.,][]{2021PhRvL.127e5101B,Selvi2024}. Most critically, the rotation profiles of LGRB progenitors remain poorly constrained, leaving open the question of whether a centrifugally supported disk forms during the PNS phase \citep[see, e.g.,][]{Gottlieb_2024,2024PhRvD.109d3051S,Fryer2025}. If no disk exists prior to the PNS collapse, the external currents needed to anchor the magnetic field lines vanish, preventing the BH from retaining the PNS magnetic flux.

An alternative to the PNS inheritance scenario is that the large-scale poloidal magnetic field is generated in situ within the collapsar disk through some dynamo processes \citep{1996ApJ...464..690H}. In particular, the dynamo driven by the magnetorotational instability (MRI) \citep{brandenburg_dynamo-generated_1995,rincon_self-sustaining_2007,rincon_subcritical_2008,riols_global_2013,riols_magnetorotational_2017,gressel_characterizing_2015,mamatsashvili_zero_2020,2022MNRAS.516.4346G,held_mri_2024,2024MNRAS.532.1522J} has been shown to self-consistently amplify and generate the large-scale poloidal field required for jet launching in a variety of systems, including steady-state accreting tori \citep{2019MNRAS.490.4811C,2020MNRAS.494.3656L,2024ApJ...960...97R}, post-binary neutron-star merger disks \citep{Palenzuela2022,2023ApJ...954L..21G,Hayashi2023,Izquierdo2025,Gutierrez2026}, and the accretion-induced collapse of (magnetized) white dwarfs \citep{Combi2025}.

In most previous demonstrations of large-scale poloidal magnetic field generation, the accretion flow begins with significant magnetization (plasma $\beta^{-1} \sim 0.2$), either imposed artificially in steady-state tori or arising naturally from strong shear and differential rotation in post-merger disks and accretion-induced collapse. Collapsar disks face two major disadvantages compared to other astrophysical environments. First, they sustain high mass accretion rates over extended periods due to continuous replenishment of infalling material, making it more difficult to build up significant disk magnetization\footnote{Even if a jet can be successfully produced, it may still be significantly perturbed or shut down by the ram pressure of freely falling gas; see \citet{2025ApJ...985..135C}.}. Second, their seed magnetic fields are weak ($\sim 10^{12}$\,G) compared to the jet-launching criteria \citep[$\sim 10^{15}$\,G, see][]{2022MNRAS.510.4962G, 2023ApJ...952L..32G}. Together, these factors leave open the question of whether a dynamo can operate efficiently in collapsars.

Recently, \citet{2025PhRvD.111l3017S} presented the first attempt to model dynamo action in collapsars using axisymmetric, general-relativistic magnetohydrodynamics (GRMHD) simulations. However, because large-scale dynamos cannot be sustained self-consistently under axisymmetry, they implemented the dynamo through parameterized prescriptions. Owing to the complexity of modeling collapsar systems, requiring long integration times, a large computational domain to capture the infalling gas, and sufficiently high resolution to resolve (M)HD instabilities, 3D simulations have thus far remained computationally prohibitive.

In this letter, we present the first 3D GRMHD simulations of collapsars that employ physically motivated initial magnetic-field configurations --- a weak toroidal field that closely follows that of the pre-collapse stellar models. We find that an accretion-disk dynamo, characterized by strong azimuthal shear, operates self-consistently, generating large-scale, strong, and coherent poloidal magnetic fields that power relativistic jets. This demonstrates, for the first time, the self-consistent formation of collapsar dynamo-driven jets. The paper is organized as follows. Section~\ref{sec:method} describes the simulation methodology, including the construction of the initial conditions, the pre-BH formation dynamical evolution, and the grid setup. Section~\ref{sec:mhdevolve} presents the dynamics of post-BH formation, and Section \ref{sec:dynamo} discusses the dynamo processes in the collapsar disk. Section~\ref{sec:conclude} discusses the implications of our findings, compares them with previous studies, and summarizes our conclusions.

\section{Numerical Simulations} \label{sec:method}
\begin{figure*}[htb!]
    \centering
    \gridline{
    \fig{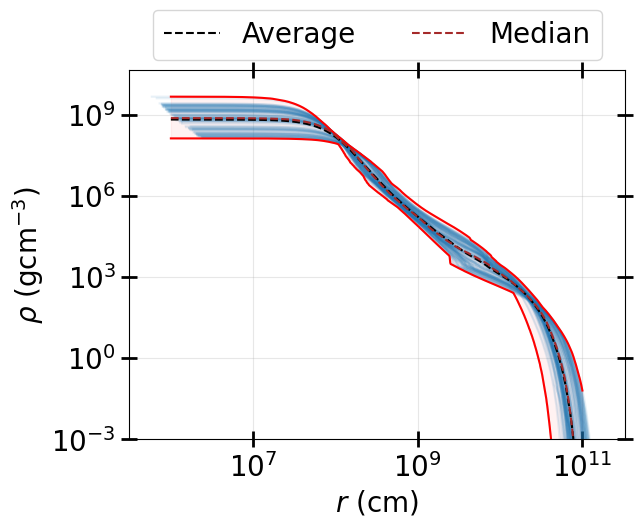}{0.5\textwidth}{(a)}
    \fig{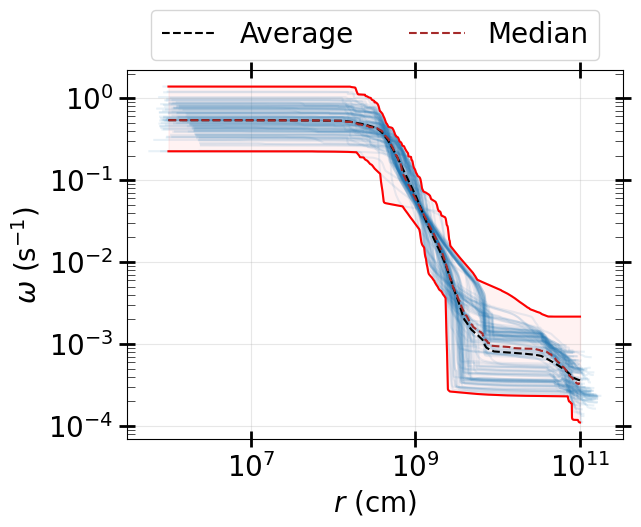}{0.5\textwidth}{(b)}}
    \gridline{
    \fig{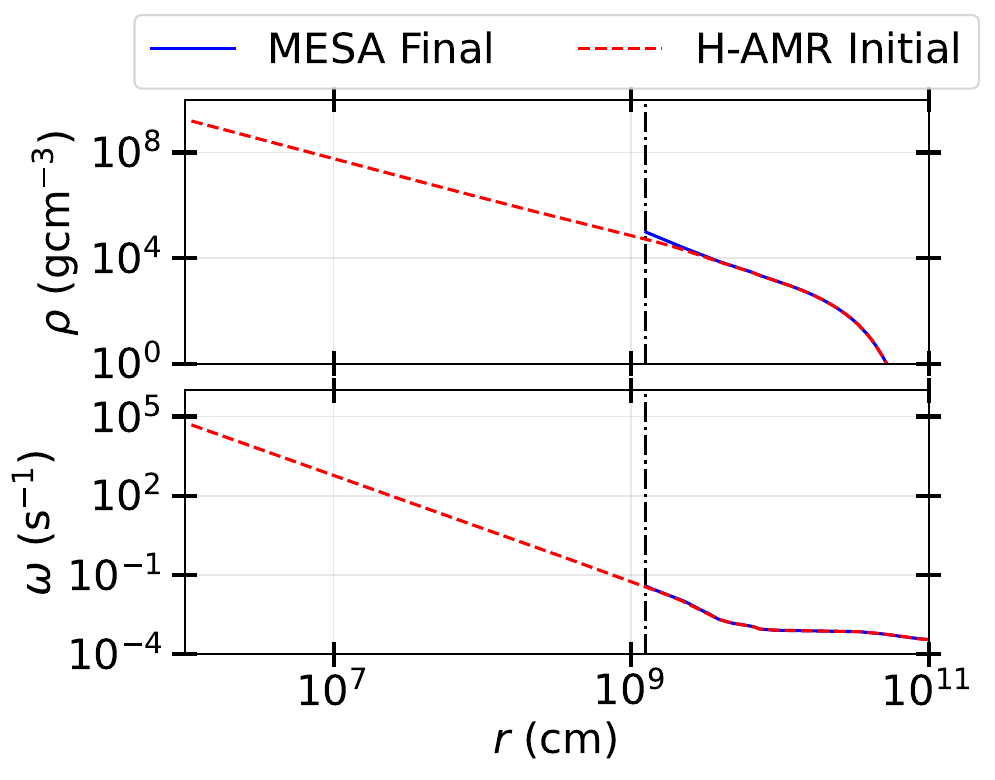}{0.58\textwidth}{(c)}
    \fig{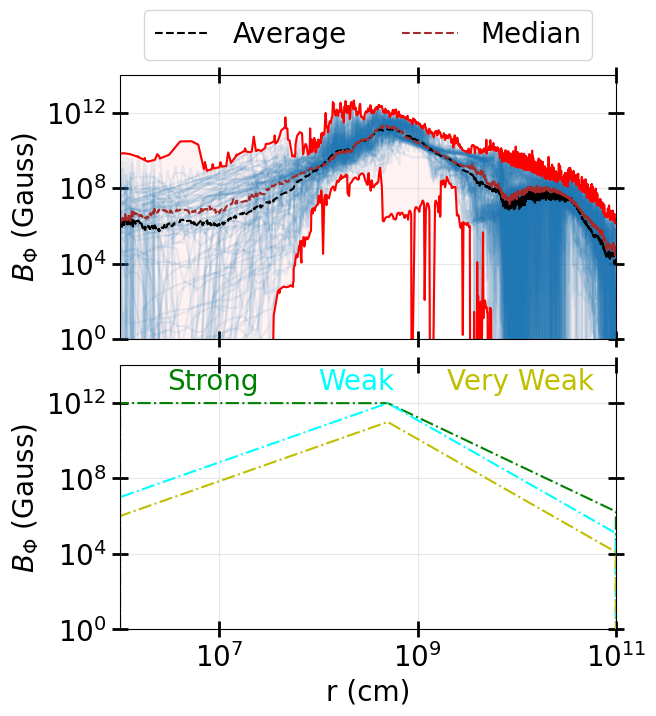}{0.42\textwidth}{(d)}}
    \caption{Radial profiles of density, angular momentum, and magnetic fields across 120 \texttt{MESA} models. In panel (a), we show the fluid density, and in panel (b), the angular velocity: The blue curves display the radial profiles of the different \texttt{MESA} models. The upper and lower red curves, which bracket the blue curves, represent the maximum and minimum values at each radius across all models. The black dashed line denotes the mean profile, and the brown dashed line shows the median. Panel (c) shows the mean density profile (upper panel) and angular velocity profile (lower panel), with the inner $3\,M_{\odot}$ of fluid removed (solid blue line) as it forms a BH. The red dashed curve shows the profile of the gas outside the enclosed BH mass coordinate as it free-falls toward the BH horizon. The vertical black dash-dotted line marks the radial coordinate enclosing $3$\,$M_{\odot}$ at the time of core collapse. Panel (d) exhibits the same information as in (a) and (b), but for the toroidal magnetic field. When averaging the toroidal magnetic field, we exclude all data points with field strengths below $1$\,G from the statistics. In the same plot, we append the initial magnetic field profiles corresponding to the Strong (green), Weak (cyan), and Very Weak (yellow) cases. We separate the raw data from the analytic fit for clearer visualization. \label{fig:mesa_initial_combine}}
\end{figure*}
We initialize the simulations with a spherically symmetric star surrounding a newborn BH of mass $M =3\,M_{\odot}$\footnote{The maximum mass of a neutron star is uncertain and depends on both the nuclear equation of state and the rotational profile \citep{1974PhRvL..32..324R, 1996ApJ...470L..61K, 2000ApJ...528L..29B, 2004ApJ...610..941M}, so we adopt a fiducial value of $3\,M_\odot$.}, and set its natal dimensionless spin to be close to the maximum equilibrium PNS spin, where we adopt a value of $a = +0.5$ \citep[see, for instance, Equation 14 of][]{Gottlieb_2024}. We evolve the plasma dynamics by solving the ideal GRMHD equations with the GPU‑accelerated code \texttt{H‑AMR} \citep{2022ApJS..263...26L}, and provide the full set of equations in Appendix~\ref{sec:grmhd}. Because \texttt{H-AMR} employs a static metric, the BH mass and spin remain fixed throughout the simulation. We model the accreting plasma as an ideal gas with an adiabatic index of $\gamma = 4/3$, representative of radiation-dominated gas, as expected for the massive progenitors considered here.

The radial profiles of the rest-mass density, internal energy, and magnetic field are taken from \texttt{MESA} models computed with the same setup as in \citet{Gottlieb_2024}. In those models, angular-momentum transport in radiative regions is provided by magnetic torques via the TSD \citep{2002A&A...381..923S}, while angular momentum is transported diffusively in convective layers. Mass loss via winds is modeled as line‑driven \citep{2000A&A...362..295V, 2001A&A...369..574V}, and the model adopts the empirical Wolf-Rayet wind prescription of \citet{2000A&A...360..227N} when the surface helium mass fraction falls below $0.4$. The progenitor models span zero-age main-sequence masses of $(30 - 96)\,M_{\odot}$ and initial rotation rates of $(0.5 - 0.99)$ times the critical angular velocity, $\omega_{\rm crit}=\sqrt{(1-L/L_{\rm Edd})GM_{*}/R_{*}^{3}}$ with $L/L_{\rm Edd}$ the ratio of the luminosity to the Eddington luminosity, $M_{*}$ stellar mass, and $R_{*}$ equatorial radius of the star. A $\omega\ge\omega_{\rm crit}$ frequency would result in centrifugal and radiative forces exceeding the gravitational forces at the stellar equator.

Importantly, for our focus on the collapse dynamics, these models were computed with a 128-isotope nuclear reaction network (\texttt{mesa\_128.net}). This allows us to follow self-consistently the deleptonization of the core \cite[e.g.,][]{1977ApJS...35..145A, 1996ApJ...460..869H}, as opposed to small ($\sim{}$20-isotope) reaction networks that predetermine the electron-fraction \citep{2000ApJS..129..377T}, the effective Chandrasekhar mass of the cores, and ultimately the pressure, density, and angular momentum profile at the onset of collapse (e.g., \citealt{2016ApJS..227...22F} and \citealt{2024RNAAS...8..152R} for rotating progenitors). This constitutes an important improvement on the progenitor structures compared to previous studies. We define core collapse as the moment when the infall velocity in the Fe core reaches $300\,\mathrm{km\,s^{-1}}$, sufficiently late that the Alfv\`en timescale is typically shorter than the remaining time to core-bounce \citep{Gottlieb_2024}. At the onset of core collapse, our grid yields minimum and maximum stellar masses of $19.5\,M_{\odot}$ and $45.7\, M_{\odot}$, with corresponding radii of $1.48\,R_{\odot}$ and $1.60\,R_{\odot}$, respectively.

To determine representative \texttt{MESA} profiles, we compute an ensemble average across all $120$ models and perform a semi-analytical fit to the aforementioned quantities. Since the simulation does not include self-gravity to hold the gas together, we set the initial gas internal energy profile to be $u_g (r) =10^{-6}\,GM\rho (r)/r$. Figure~\ref{fig:mesa_initial_combine} (a), (b) depict the mass density and angular velocity profiles of all \texttt{MESA} models (blue), respectively. The ensemble-averaged profiles (dashed black lines) are close to the median profiles (dashed brown lines) for our initial conditions. We note that the average profiles do not represent any astrophysically meaningful weighting (e.g., the initial mass function or the initial rotation distribution).

We do not map the \texttt{MESA} profiles directly into \texttt{H-AMR} because the inner $3$\,$M_{\odot}$ of the star collapses into a BH. Thus, we remove all the plasma within the BH-forming radius, $r_{\rm BH}$, which encloses $3$\,$M_{\odot}$. To reduce computational cost, we semi-analytically evolve the gas outside the $3\,M_\odot $ mass coordinate. A detailed description of the procedure for the density profile is provided in Appendix~\ref{sec:semi-analytic}. The top panel of Figure~\ref{fig:mesa_initial_combine} (c) shows the semi-analytically evolved density profile at the moment when  the gas reaches the BH horizon (red dashed line). The gas profile exhibits the characteristic free-fall $r^{-3/2}$ power law, whereas the outer parts remain largely unchanged, since their free-fall time is much longer than that of the gas that reached the horizon. Our fit to the radial density profile at the time of BH formation is
\begin{equation}
    \rho (r) \propto r^{-1.5}\left( 1 - \frac{r}{R_*} \right)^{5.7},
\end{equation}
where $R_*=10^{11}$\,$\mathrm{cm}$ is the stellar radius. The proportionality constant is chosen such that the total mass within the computational domain is $20$\,$M_{\odot}$. Given this density profile, the stellar binding energy is $\sim 10^{52}\,\mathrm{erg}$. For comparison, the total magnetic energy at the end of the simulations for the jetted model reaches a few $\times 10^{52}\,\mathrm{erg}$.

The bottom panel of Figure~\ref{fig:mesa_initial_combine} (c) shows the semi-analytically evolved angular velocity profile, which follows an $r^{-2}$ scaling in the inner region due to angular momentum conservation, then flattens and finally transitions into a second power law decline. Accordingly, we model the angular velocity profile using a double-broken power law.
\begin{equation} \label{eqn:omega}
    \omega(r) = \begin{cases}
    j_{\rm shell}/r^2, & \text{if } r \leq r_{1j}\\
    j_{\rm shell}/r_{1j}^{2}, & \text{if } r_{1j} \leq r \leq r_{2j}\\
    j_{\rm shell}(r_{2j}/r_{1j})^{2}/r^{2}, & \text{otherwise, }
    \end{cases}
\end{equation}
where $j_{\mathrm{shell}}$ denotes the specific angular momentum at the enclosed mass coordinate of $3$\,$M_{\odot}$, $r_{1j}=9.1\times 10^9$\,$\mathrm{cm}$ and $r_{2j}=5.8\times 10^{10}$\,$\mathrm{cm}$.

Figure~\ref{fig:mesa_initial_combine} (d) depicts the toroidal magnetic field profiles of all models (blue). The ensemble averaging is performed in log$_{10}$ space of the magnetic field strength, and excludes regions where the field is weaker than $1$\,G. We fit the average toroidal field profile, which aligns with the median, with a broken power law model
\begin{equation} \label{eqn:bfield}
    B^{\phi}(r) = \begin{cases}
    B_0\,(r/r_{1B})^{\alpha}, & \text{if } r_{0B} \leq r \leq r_{1B} \\
    B_0\,(r_{1B}/r)^{\beta}, & \text{if } r_{1B} \leq r \leq r_{2B} \\
    0, & \text{otherwise, }
    \end{cases}
\end{equation}
where $r_{0B}=10^6$\,cm, $r_{1B}=5\times 10^8$\,cm, and $r_{2B}=10^{11}$\,cm. We note that we do not evolve the toroidal field alongside the other quantities under the assumption of free fall. This limitation arises from our lack of knowledge about how the toroidal field should behave under compression (flux-freezing). We also do not consider the poloidal field in this study because it is much weaker than the toroidal component and is largely canceled out during collapse \citep{Gottlieb_2024}.

To examine the effects of the disk size and seed magnetic field on the dynamo process, we set our free parameters to be $j_{\mathrm{shell}}$, $B_0$, $\alpha$, and $\beta$, as summarized in Table \ref{tab:models}. We vary $j_{\mathrm{shell}}$ across $5.63\times 10^{16}$, $8\times 10^{16}$, and $1.13\times 10^{17}$\,cm$^2$\,s$^{-1}$ to examine gas rotating at different speeds. This corresponds to rotating shells with different circularization radii, $r_{\mathrm{circ}}$: $18$, $36$, and $72$\,$GM/c^2$, respectively, where $GM/c^2 = 4.43\times10^{5}\,{\rm cm}$ is the BH gravitational radius. We define a Strong toroidal-field model with $B_0=10^{12}$\,G, $\alpha = 0.0$, and $\beta = 2.5$; a Weak model with $B_0=10^{12}$\,G, $\alpha = 1.85$, and $\beta = 3.0$; and a Very Weak model with $B_0=10^{11}$\,G, $\alpha = 1.85$, and $\beta = 3.0$. The Strong, Weak, and Very Weak profiles are shown in the bottom panel of Figure~\ref{fig:mesa_initial_combine} (d) as green, cyan, and yellow lines, respectively. The maximum plasma parameter $\sigma =b^2/\rho$ in the Strong field case is $\sigma \sim10^{-4}$.

\begin{deluxetable}{ccccc}[htb!]
\setlength{\tabcolsep}{1pt}
\tablecaption{List of simulations and model parameters. The first column lists the model names following the circularization radius, $r_{\mathrm{circ}}$, and the field strength. $j_{\mathrm{shell}}$ is the specific angular momentum at the initial enclosed mass coordinate of $M(r)=3$\,$M_{\odot}$, which corresponds to $r_{\mathrm{circ}}$. The magnetic-field models, Strong, Weak, and Very Weak, are distinguished by the peak initial toroidal field strength and the power law indices of their radial variations. $T_f$ denotes the simulation time in seconds. \label{tab:models}}
\tablewidth{0pt}
\tablehead{
\colhead{Model} & \colhead{$j_{\rm shell}\,\left[{\rm cm}^{2}\,{\rm s}^{-1}\right]$} & \colhead{$r_{\rm circ}$ $\left[10^{7}\,\mathrm{cm}\right]$} & \colhead{$B$-Field} & \colhead{$T_f$ [s]}}
\startdata
\texttt{rc\_18\_B\_s} & $5.6\times10^{16}$ & $0.80$ & Strong & $4.0$ \\
\texttt{rc\_36\_B\_vw} & $8\times10^{16}$ & $1.59$ & Very Weak & $4.0$ \\
\texttt{rc\_36\_B\_w} & $8\times10^{16}$ & $1.59$ & Weak & $4.3$ \\
\texttt{rc\_36\_B\_s} & $8\times10^{16}$ & $1.59$ & Strong & $3.0$ \\
\texttt{rc\_72\_B\_s} & $1.1\times10^{17}$ & $3.19$ & Strong & $4.0$
\enddata
\end{deluxetable}

The computational domain is in spherical polar coordinates, spanning the full range of polar ($\theta$) and azimuthal ($\phi$) angles. The inner radial boundary is located at $r = 1.16$\,$GM/c^2$, placing it $5$ cells inside the event horizon, while the outer radial boundary is set at $r = 10^{5}$\,$GM/c^2$. The radial grid is uniform in logarithmic coordinates, so that $\delta r/r$ is constant, with $\delta r$ being the cell size in the radial coordinate. We employ static mesh refinement (SMR) to resolve the computational domain near the BH. The base resolution is $224\times 144\times 128$, which yields a cell-size ratio of $r:r\,\delta\theta:r\,\delta\phi \approx 2.33:1:2.25$. We use two levels of refinement within the region $4\, GM/c^2\leq r\leq 500\, GM/c^2$, to ensure that the MRI, which is the relevant instability in accretion disks, is resolved (see Appendix~\ref{sec:resolution} for the mesh-block distribution at the highest refinement level and resolution convergence of the MRI). This yields a total effective number of $896\times 576\times 512$ grid cells. The innermost stable circular orbit (ISCO) of a BH with spin $a = +0.5$ is located at $4.2\,GM/c^2$. We additionally apply four levels of internal and external de-refinement to prevent excessive grid density near the pole. External de-refinement reduces the number of blocks in the base layer within $30^{\circ }$ of the pole, whereas internal de-refinement spatially averages the cells' fluxes and conserved quantities across the $\phi$ direction. See Section 3.4 of \citet{2022ApJS..263...26L} for details.

\section{Jet Dynamics} \label{sec:mhdevolve}

\subsection{Jet Launching} \label{subsec:launch}
\begin{figure*}[htb!]
    \centering
    \includegraphics[width=1.0\textwidth]{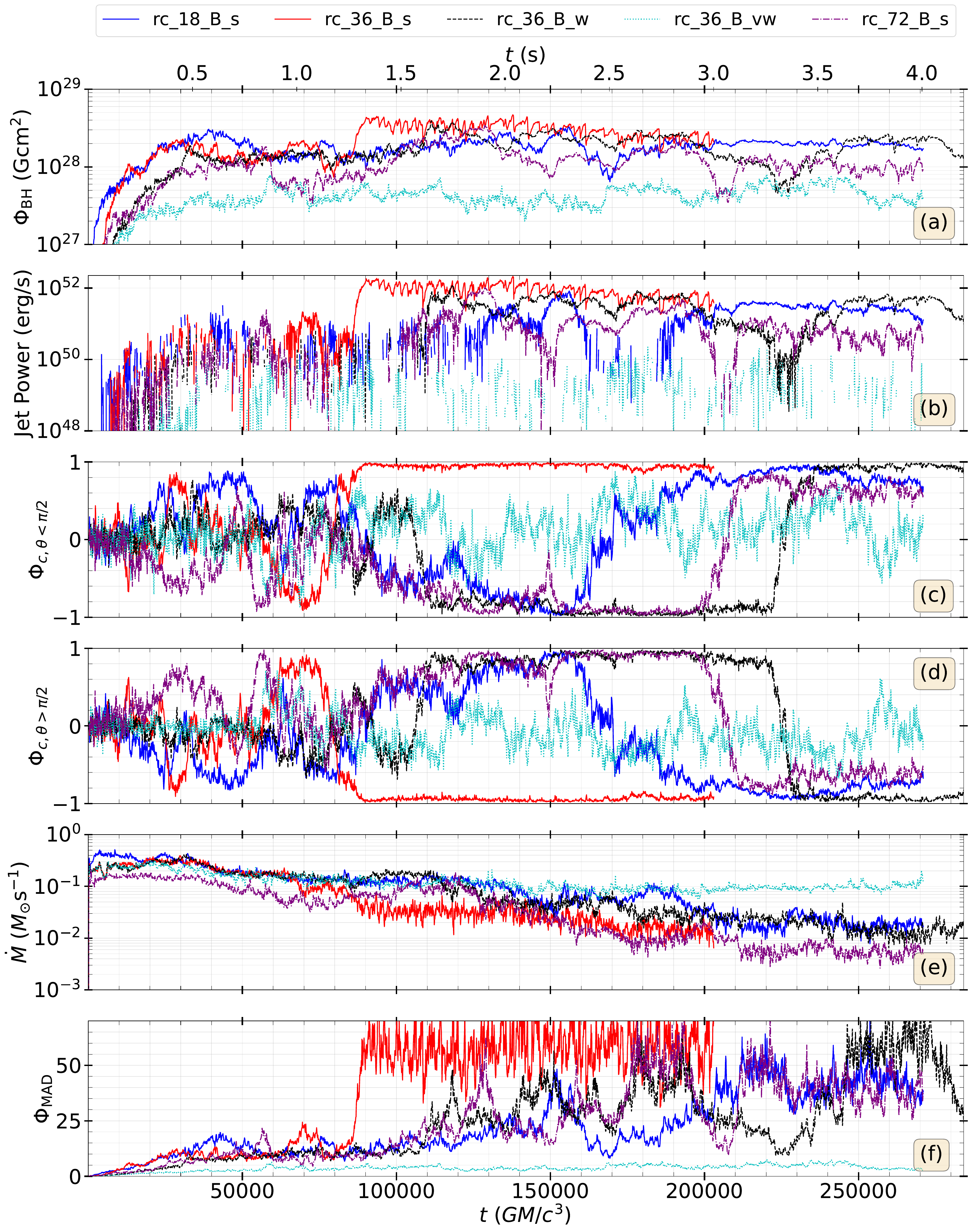}
    \caption{Timeseries of all models showing: horizon-penetrating magnetic flux (a), jet power (b), upper-hemisphere magnetic-field coherency (c), lower-hemisphere magnetic-field coherency (d), mass accretion rate (e), and the dimensionless horizon-penetrating magnetic flux, i.e., the MAD parameter (f). \label{fig:timeseries}}
\end{figure*}

Figure~\ref{fig:timeseries} summarizes the timeseries diagnostics that describe the dynamics of the plasma near the BH in all models. Figure~\ref{fig:timeseries} (a) shows the dimensional horizon-penetrating magnetic flux
\begin{equation}
    \Phi_{\rm BH} = \sqrt{4\pi} \int_{r=r_{\rm BH}} |B^{r}|\sqrt{-g}d\theta d\phi.
\end{equation}
Here, $B^{r}$ is the radial component of the $3-$magnetic field, and $\sqrt{-g}$ is the metric determinant. The requirement that the magnetic energy density near the BH exceed the kinetic energy density of the accreting plasma in order to launch a jet \citep{2009MNRAS.397.1153K,2023ApJ...952L..32G} places a stringent lower bound on the magnetic flux threading the BH horizon in LGRBs. \citet{2022MNRAS.510.4962G, 2023ApJ...952L..32G} showed that for $\dot{M} \sim \mathcal{O}(0.1)\,M_{\odot}$\,s$^{-1}$, this threshold translates to $\Phi_{\rm BH} \sim\mathcal{O}(10^{28})$\,G\,cm$^2$. Our results indicate that all models, except \texttt{rc\_36\_B\_vw}, reach fluxes of this order within $\sim0.5$\,s. In contrast, model \texttt{rc\_36\_B\_vw} attains only $\Phi_{\mathrm{BH}} \sim \mathcal{O}(10^{27})$\,G\,cm${^2}$ and remains at that level.

Figure~\ref{fig:timeseries} (b) shows the BZ jet power, computed as
\begin{equation}
    L_{\rm BZ} = \int_{r=r_{\rm BH}, \sigma > 1} (-T^{r}_{t} - \rho u^{r})\sqrt{-g}d\theta d\phi,
\end{equation}
where $T^{\mu}{}_{\nu}$ is the stress–energy tensor and $u^{r}$ is the radial four–velocity. The integral is evaluated over cells at the event horizon with $\sigma > 1$. The jet power fluctuates between positive and values (i.e., accretion-dominated) during the first $t \gtrsim 1$\,s, after which, at $\sim 1.2 - 2$\,s, a steady jet power is established. The first and most robust jet-launching event occurs in model \texttt{rc\_36\_B\_s} at $\sim 1.25 - 1.3$\,s, coinciding with an increase in $\Phi_{\mathrm{BH}}$. The BZ luminosity reaches $\sim 10^{52}\,\mathrm{erg\,s^{-1}}$, after which it declines on the long run, while exhibiting cycles of gradual buildup and rapid decay on smaller timescales, indicating that the system enters the magnetically arrested (MAD) state (see the discussion below). The latest onset of jet production is seen in model \texttt{rc\_18\_B\_s} at $\sim 2 - 2.5$\,s. Models \texttt{rc\_18\_B\_s}, \texttt{rc\_36\_B\_w}, and \texttt{rc\_72\_B\_s} display strongly fluctuating jet power around $10^{51}\,\mathrm{erg\,s^{-1}}$, with the first and third models transitioning to a steady jet-launching state with $L_{\mathrm{BZ}} \sim 2\times 10^{51}\,\mathrm{erg\,s^{-1}}$ at $\sim 3\,\mathrm{s}$. Model \texttt{rc\_36\_B\_vw}, which does not show jet activity\footnote{Although the MRI quality factor in model \texttt{rc\_36\_B\_vw} exceeds the standard criterion for resolving MRI, it is considerably lower than other models (Figure~\ref{fig:supp}), potentially leading to an inability to produce large-scale, strong poloidal fields.}, exhibits an energy outflow rate (measured at the event horizon) below $10^{50}\, \mathrm{erg\, s^{-1}}$.

Sufficient magnetic flux can thread the BH without producing a jet in the absence of magnetic-field coherence, which can be estimated by \citep{2024ApJ...960...97R, santhiya2025dynamojetinterconnectionsgrmhd}
\begin{equation}
  \Phi_{c, \theta < \pi/2} = \frac{\int_{r = r_{\rm BH}, \theta < \pi/2} B^r \sqrt{-g}d\theta d\phi}{\int_{r = r_{\rm BH}, \theta < \pi/2} |B^r| \sqrt{-g}d\theta d\phi}.
\end{equation}
Although this definition applies to the upper hemisphere of the BH, it can be trivially generalized to measure the coherence in the lower hemisphere as well. Figure~\ref{fig:timeseries} (c), (d) demonstrate that in all models $|\Phi_c|$ fluctuates around zero at early times. A jet emerges only once $|\Phi_c|$ reaches order unity. In particular, model \texttt{rc\_36\_B\_s} reaches $|\Phi _c|\sim 1$ at $\sim 1.25 - 1.3$\,s, after which $\Phi_c$ does not switch sign. Interestingly, models \texttt{rc\_18\_B\_s}, \texttt{rc\_36\_B\_w}, and \texttt{rc\_72\_B\_s} also attain $|\Phi _c|\sim 1$ and launch a jet for a transient period, but subsequently $\Phi _c$ reverses sign and then reaches $|\Phi _c| \sim 1$ at later times. The times at which $\Phi _c$ switches sign coincide with drops in both $\Phi _{\mathrm{BH}}$ and $L_{\mathrm{BZ}}$. This behavior indicates a magnetic flux inversion \citep{2024MNRAS.532.1522J, 2025ApJ...979..248K} and the formation of a striped jet, which will be discussed further in Section~\ref{subsec:jetvar}. Model \texttt{rc\_36\_B\_vw} never attains $|\Phi _c|\sim 1$; instead, its value fluctuates around zero, consistent with a lack of jet activity.

Figure~\ref{fig:timeseries} (e) depicts the mass accretion rate onto the BH, which is calculated as
\begin{equation}
  \dot{M} = \int_{r = r_{\rm BH}}\rho u^r\sqrt{-g}d\theta d\phi.
\end{equation}
Model \texttt{rc\_36\_B\_vw} maintains a mass accretion rate of $\dot{M}\gtrsim 10^{-1}\, M_{\odot }\, \mathrm{s^{-1}}$, remaining at this value throughout the evolution. This behavior is attributed to the absence of a large-scale poloidal field capable of producing a jet, which would otherwise impede the inflow. In all other models, $\dot{M}$ steadily decreases from $\sim 10^{-1}\, M_{\odot }\, \mathrm{s^{-1}}$ to $\sim 10^{-2}\, M_{\odot }\, \mathrm{s^{-1}}$ due to the ongoing jet activity. For the duration of our simulations, the estimated BH mass growth is $\lesssim 0.1\,M_\odot$, justifying the fixed BH mass and spin approximation \citep[see][]{2024ApJ...961..212J}.

The accumulation of magnetic flux near the BH renders the disk magnetically arrested \citep[MAD;][]{2011MNRAS.418L..79T}, oversaturating the BH with magnetic flux \citep{2003PASJ...55L..69N,2008ApJ...677..317I,2012MNRAS.423.3083M}. This process can be quantified by the MAD parameter
\begin{equation} \label{eqn:mad}
\Phi_{\mathrm{MAD}}= \frac{\Phi_{\mathrm{BH}}c^{3/2}}{\dot{M}^{1/2}GM}.
\end{equation}
Figure~\ref{fig:timeseries} (f) demonstrates that the model without jets, \texttt{rc\_36\_B\_vw}, maintains $\Phi _{\mathrm{MAD}}<5$ throughout, below the critical $\Phi _{\mathrm{MAD}}\approx 10$ required for jet launching \citep{2023ApJ...952L..32G,2025ApJ...985L..26I}. In contrast, model \texttt{rc\_36\_B\_s} exhibits a sharp increase in the MAD parameter from $\sim 15$ to $\gtrsim 60$ at $\sim 1.25 - 1.3$\,s, after which it saturates. The resulting value is consistent with the saturated MAD level expected for a BH with spin $a = +0.5$ \citep{2022MNRAS.511.3795N}, indicating that the system has entered the MAD state. The fluctuating pattern in $\Phi _{\mathrm{BH}}$ is attributed to magnetic-flux eruptions in a MAD disk, driven by a reconnecting current sheet near the BH \citep{2020MNRAS.497.4999D, 2022ApJ...924L..32R, jacqueminide2025demystifyingfluxeruptionsmagnetic}.

The steady increase in $\Phi_{\mathrm{MAD}}$ correlates with the growth in coherency of $\Phi_{c,\theta}$ (in model \texttt{rc\_18\_B\_s} at $\sim 2.5 - 3$\,s, model \texttt{rc\_36\_B\_s} at $\sim 1.25 - 1.3$\,s, model \texttt{rc\_36\_B\_w} at $\sim 3 - 3.5$\,s, and model \texttt{rc\_72\_B\_s} at $\sim 3 - 3.5$\,s). However, model \texttt{rc\_36\_B\_w} attains $\Phi _{\mathrm{MAD}}\sim 60$ between $\sim 3.5$ and $4.0$\,s, before its value declines alongside $\Phi_{\mathrm{BH}}$. As we shall show, this decline arises from the advection of reversed poloidal fields that are dragged onto the BH and reconnect with the pre‑existing poloidal field.

\subsection{Variability and Morphology} \label{subsec:jetvar}
\begin{figure*}[htb!]
    \centering
    \includegraphics[width=1.0\textwidth]{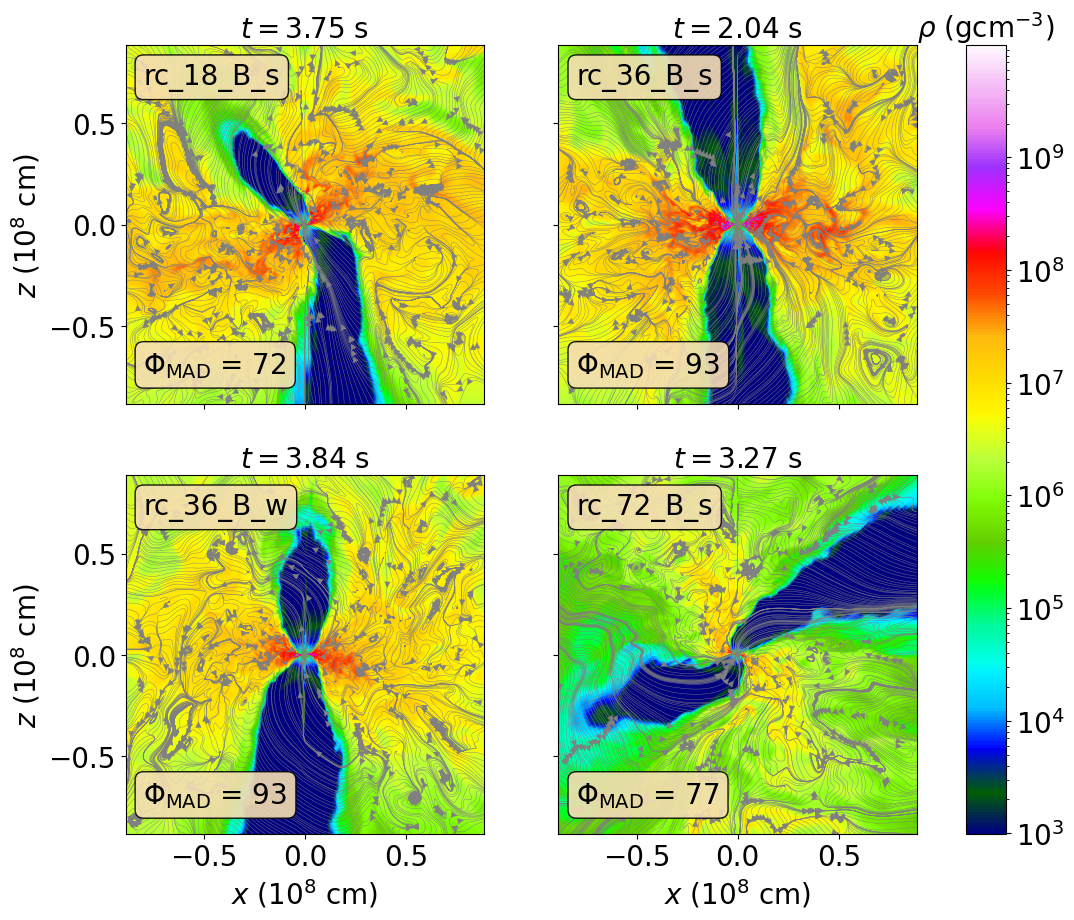}
    \caption{Density maps (with magnetic field lines overlaid) in the $x-z$ plane for all models (except \texttt{rc\_36\_B\_vw}) at the epoch when the instantaneous MAD parameter reaches its maximum. \label{fig:density_jet}}
\end{figure*}
\begin{figure*}[htb!]
    \centering
    \includegraphics[width=1.0\textwidth]{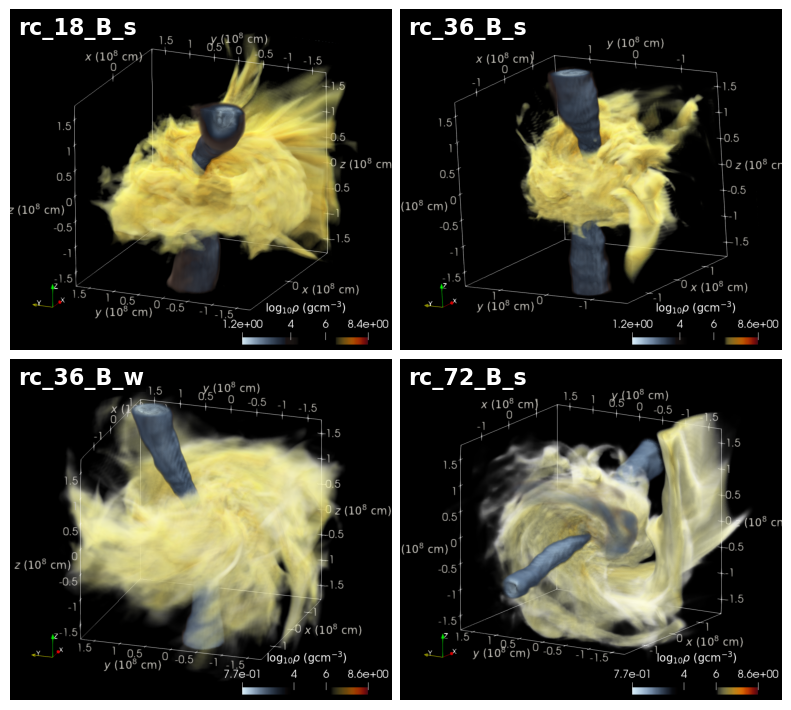}
    \caption{Same as Figure~\ref{fig:density_jet}, but showing the 3D volume rendering of the rest-mass density on a log$_{10}$ scale, illustrating wobbling jet morphologies in all models. The lower‑left arrows indicate the orientation. The range of densities not shown is indicated as transparent along the colorbar.\label{fig:merged_figure_collapsar}}
\end{figure*}

Most of our models exhibit flipping of the magnetic coherence parameter, suggesting that the jet switches on and off and that its toroidal field direction changes stochastically, i.e., a striped jet \citep{2002A&A...391.1141D, 2015MNRAS.446L..61P, 2019MNRAS.484.1378G, 2020MNRAS.494.4203M, 2021MNRAS.508.1241C, 2023ApJ...954...40K}. The collapsar disk is also continuously perturbed by infalling gas arriving from random directions. Such behavior contrasts with that in steady torus accretion, resulting in vigorous jet variability and non-trivial morphology that could explain the variability in LGRB light curves \citep[e.g.,][]{Sari1997,MacLachlan2013}.

Figure~\ref{fig:density_jet} depicts the meridional density maps with magnetic field lines overlaid in all models, except for model \texttt{rc\_36\_B\_vw}, which we do not discuss in Section~\ref{subsec:jetvar} and Section~\ref{sec:dynamo} as no jet activity is observed. To visualize the magnetic field within a quasi-orthonormal basis, we scale the $r$, $\theta,$ and $\phi$ components of the $3-$magnetic fields (in the coordinate basis) so that
\begin{equation} \label{eqn:vector}
    B_{\hat{r}} = \sqrt{g_{rr}}B^{r}, B_{\hat{\theta}} = \sqrt{g_{\theta\theta}}B^{\theta}, B_{\hat{\phi}} = \sqrt{g_{\phi\phi}}B^{\phi}.
\end{equation}
The snapshots are taken when the instantaneous MAD parameter reaches its maximum. We find that models \texttt{rc\_36\_B\_s} and \texttt{rc\_36\_B\_w} show jets whose base aligns with the BH’s spin axis. However, the upper jet in model \texttt{rc\_36\_B\_w} is truncated at $\sim 5\times10^{7}$\,cm as seen along the $x\mathrm{-}z$ plane, and the magnetic field lines in the upper jet are more curved than those in model \texttt{rc\_36\_B\_s}, suggesting that the jet is launched out of the $x\mathrm{-}z$ plane. This behavior is more prominent in models \texttt{rc\_18\_B\_s} and \texttt{rc\_72\_B\_s}, where the jets are tilted with respect to the midplane. Such tilting arises from the complex interplay between the jet-disk system and the free-falling material --- as the collapsar disk is continuously perturbed by the randomly oriented, freely falling gas via ram pressure \citep{2022ApJ...933L...9G}, the disk becomes misaligned with the spin axis, thereby creating poloidal magnetic field lines threading the BH at an angle that is also misaligned with the spin axis. In the simplified torus setup, on the other hand, the poloidal field threading the BH horizon has one of its footpoints anchored in the (almost planar-symmetric) accretion disk, allowing the field to open to infinity through differential rotation and extract rotational energy from the BH \citep{2005ApJ...620..889U, 2015MNRAS.446L..61P} more easily, thereby launching a powerful bipolar outflow. As we will demonstrate in Section~\ref{sec:dynamo}, the magnetic loops generated by the dynamo are not transported solely along the midplane.

Figure~\ref{fig:merged_figure_collapsar} illustrates the three-dimensional volume rendering of the jet-disk systems, corresponding to the same snapshots shown in Figure~\ref{fig:density_jet}. The volume is limited to a rectangular box of size $\sim (2\times10^{8}\,\textrm{cm})^3$. Although the jets shown in Figure~\ref{fig:density_jet} are truncated in the $x\mathrm{-}z$ plane, the 3D rendering reveals that the jets propagate to a distance of $\gtrsim 10^{8}$\,cm from the BH, with their direction pointing out of the $x–z$ plane. This wobbling motion has profound hydrodynamic and observational implications for LGRB jets \citep{2022ApJ...933L...9G}, which are beyond the scope of this paper.

\section{The Dynamo Action} \label{sec:dynamo}
\begin{figure*}[htb!]
    \centering
    \includegraphics[width=1.0\textwidth]{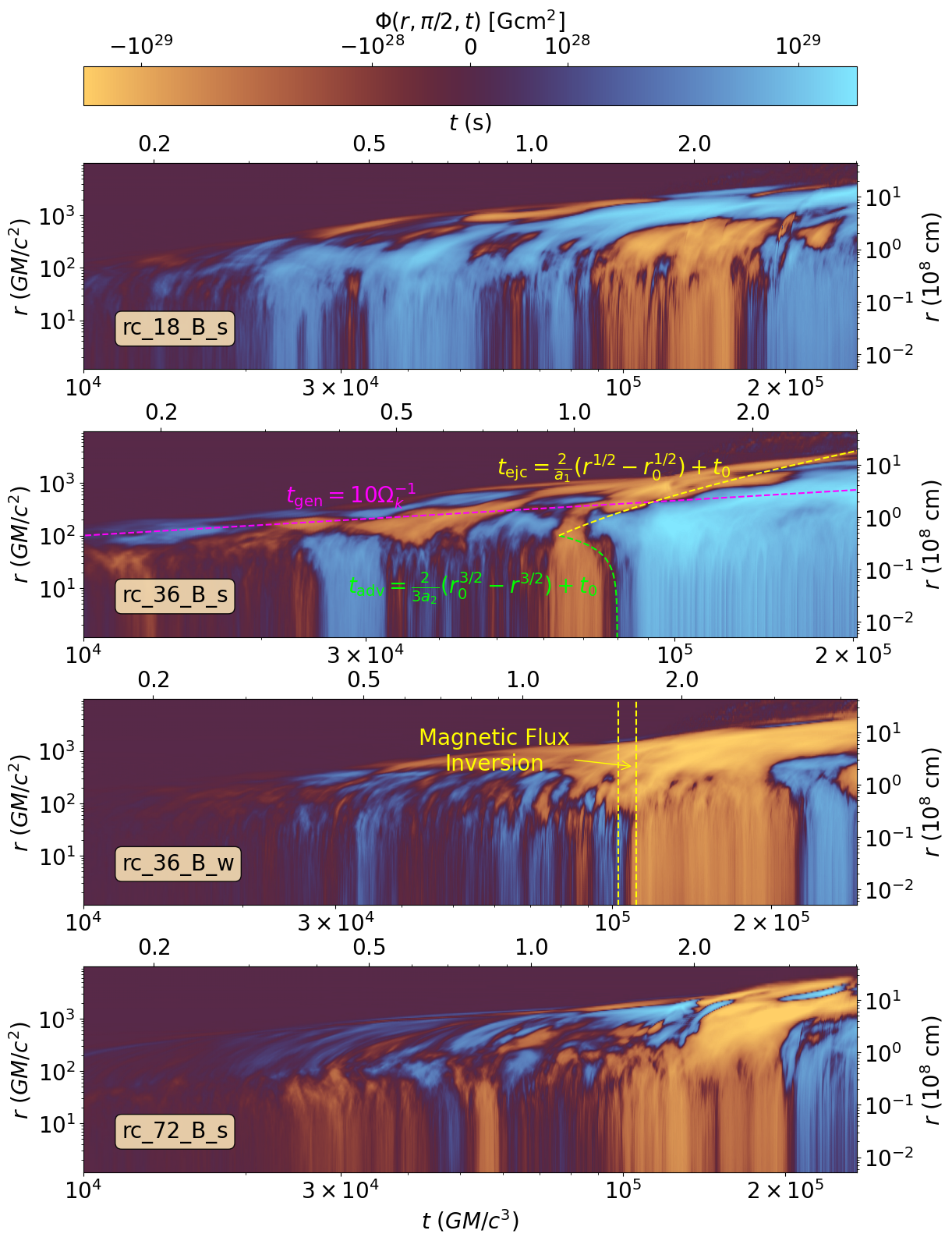}
    \caption{The spacetime diagram of the magnetic-flux function $\Phi$ evaluated on the midplane. For model \texttt{rc\_36\_B\_s}, we include the timescale for the advection, ejection, and generation of poloidal loops. For model \texttt{rc\_36\_B\_w}, we highlight the time interval during which magnetic‑flux inversion occurs without any clear signature of inward magnetic‑flux advection. We show in Appendix~\ref{sec:supp} that during this interval the magnetic loops are instead dragged inward along directions away from the midplane. \label{fig:fluxfunction}}
\end{figure*}
\begin{figure*}[htb!]
    \centering
    \gridline{
    \fig{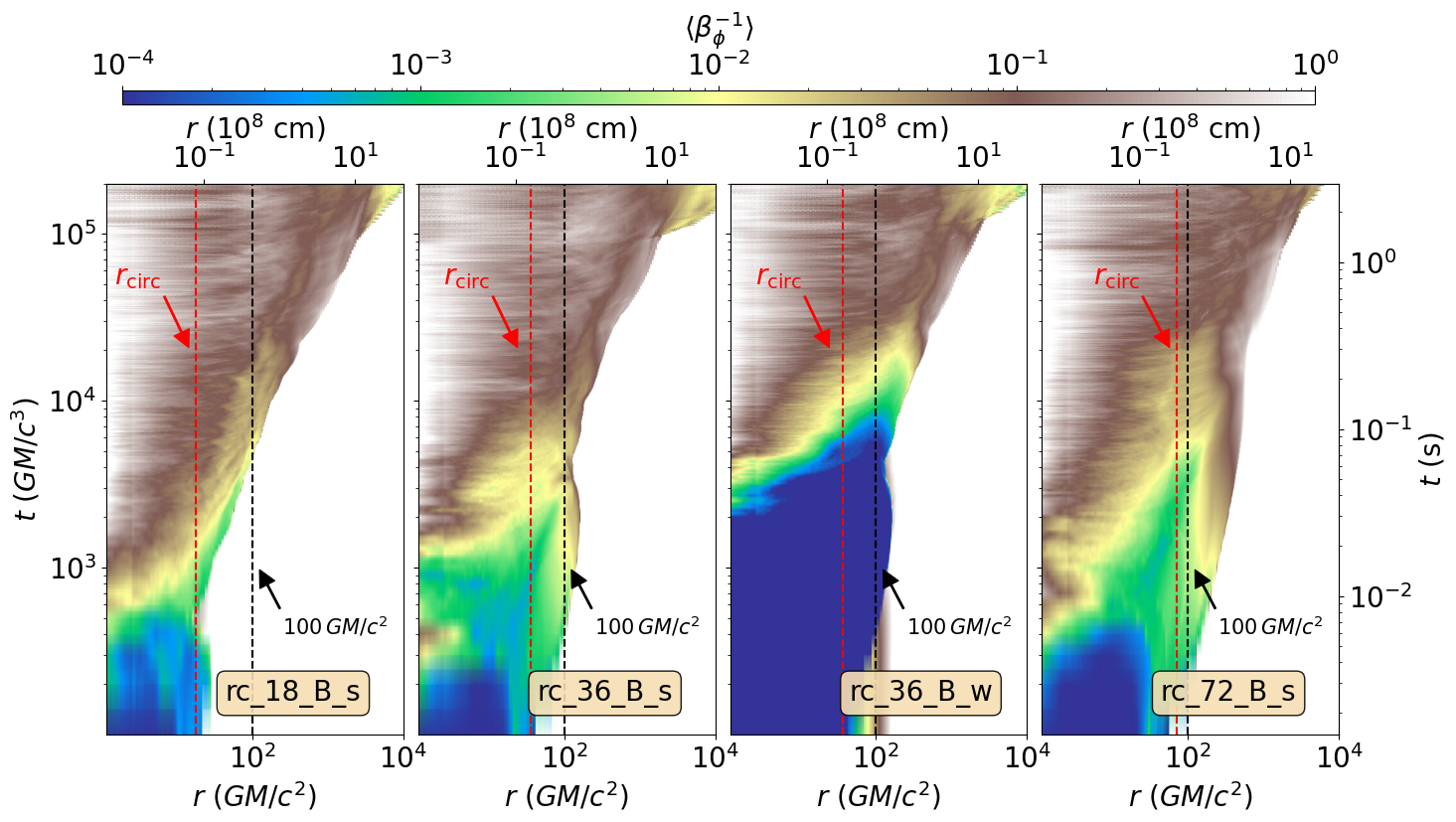}{1.0\textwidth}{(a)}}
    \gridline{
    \fig{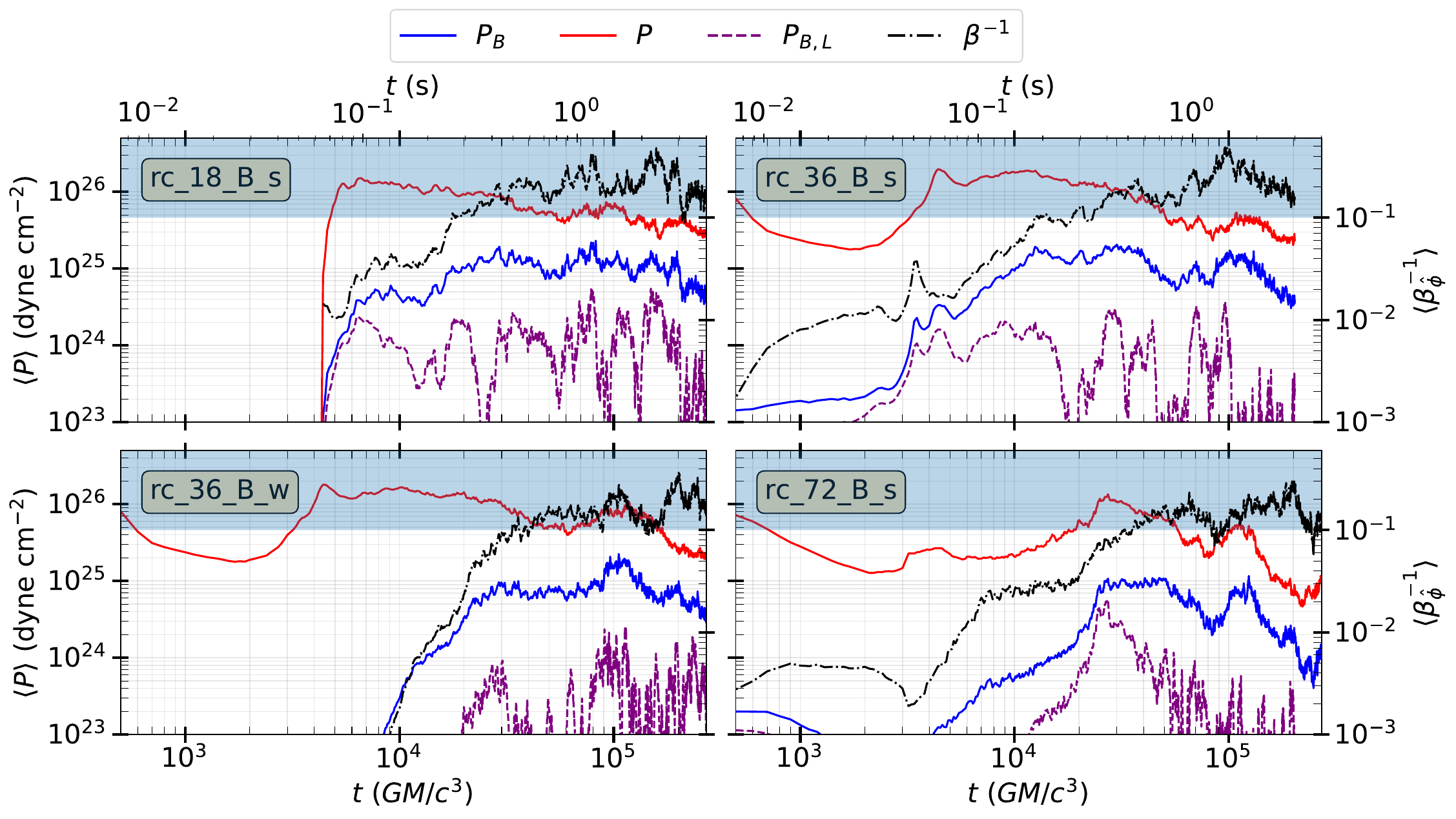}{1.0\textwidth}{(b)}}
    \caption{(a) Spacetime diagram of the mass-weighted average of the inverse toroidal plasma beta ($\langle \beta_{\hat{\phi}}^{-1} \rangle = \langle B_{\hat{\phi}}^2\rangle/[2\langle P \rangle]$). The vertical red (black) line delineates $r_{\mathrm{circ}}$ ($r = 100$\,$GM/c^2$). (b) Average toroidal magnetic pressure (blue) $\langle P_{B} \rangle = \langle B_{\hat{\phi}}^2\rangle/(8\pi)$, gas pressure ($\langle P \rangle$, red), $\langle \beta_{\hat{\phi}}^{-1} \rangle$ (black dashed), and and the large-scale averaged toroidal magnetic pressure (purple) $\langle P_{B,L} \rangle = \langle B_{\hat{\phi}} \rangle^2/(8\pi)$, all measured at $r=100\,GM/c^2$ as a function of time. The shaded region represents $\langle \beta_{\hat{\phi}}^{-1} \rangle > 0.1$. In both panels, the model name is shown in a textbox. \label{fig:dynamo_beta}}
\end{figure*}

In this section, we discuss the dynamo activity using code units, for which the natural length and time units are $ GM/c^2 = 4.43\,{\rm km}$ and $GM/c^3 = 1.48\times 10^{-5}\,\mathrm{s}$. This facilitates a more direct comparison with previous similar work \citep{2020MNRAS.494.4854D, 2024MNRAS.532.1522J}.

\subsection{Magnetic Loop Dynamics} \label{subsec:loops}
Figure~\ref{fig:fluxfunction} shows the spacetime diagram of the magnetic-flux function evaluated at the midplane, $\Phi(r, \pi/2)$, in which
\begin{equation} \label{eqn:flux_func}
  \Phi(r,\theta) = 2\pi\left(\int_{0}^{\theta} \langle B^{r}\rangle \sqrt{-g} d\theta - \int_{\pi}^{\pi - \theta} \langle B^{r}\rangle \sqrt{-g} d\theta\right).
\end{equation}
Here, $\langle B^r\rangle$ denotes the $\phi-$average of $B^r$. This diagnostic allows us to visualize poloidal magnetic loops (see Appendix~\ref{sec:supp}), as well as when and where the loops appear (on the midplane) and how they are advected toward the BH. The flux function is constructed so that each uniformly colored region corresponds to a bundle of magnetic loops sharing the same polarity, while regions of opposite color indicate loops of opposite polarity.

In model \texttt{rc\_36\_B\_s}, a magnetic loop with $ \Phi>0$ appears at a few hundred $GM/c^2$ and around $(7-8)\times 10^4$\,$GM/c^3$. The loop subsequently expands, with its leading edge being dragged toward the BH while its trailing edge expands outward. The region between the event horizon and $r \sim 1000$\,$GM/c^2$ then becomes dominated by poloidal fields of the same sign. The emergence of this dominant loop occurs just before the onset of the MAD state, as shown in Figure~\ref{fig:timeseries}. This indicates that the MAD state is triggered by the accretion of coherent, strong, dynamo‑generated poloidal magnetic loops onto the BH \citep[see also][]{2024MNRAS.532.1522J}.

We observe similar behavior in model \texttt{rc\_36\_B\_w} at $\sim 2\times 10^5$\,$GM/c^3$, but the MAD state persists for only $\sim 3\times10^{4}\,{GM/c^2}$, as another magnetic loop of comparable field strength but opposite polarity ($\Phi<0$) emerges at $\sim 2.5\times 10^5$\,$GM/c^3$, is advected inward and reconnects with the existing poloidal field, thereby shutting off the existing jet. For this model, the field loops capable of both producing and disrupting the MAD state are also generated at radii of a few hundred $GM/c^2$.

The absence of clear evidence for inward magnetic-flux advection in models \texttt{rc\_18\_B\_s} and \texttt{rc\_72\_B\_s}, and some instance of \texttt{rc\_36\_B\_w} during which magnetic-flux inversion occurs, arises from evaluating the magnetic-flux function at the midplane, which cannot capture the inward transport of magnetic loops occurring above/below the midplane in these models. As shown in Appendix~\ref{sec:supp}, magnetic loops are dragged inward along the polar region, consistent with their highly non-planar-symmetric jet and disk morphologies (see Section~\ref{subsec:jetvar}).

We find magnetic‑field loops continue to be generated with alternating polarity throughout, causing $\Phi(r,\pi/2,t)$ to oscillate between positive and negative values, except for model \texttt{rc\_36\_B\_s} which enters a persistent MAD state. This behavior reflects the stochastic nature of the dynamo and demonstrates that it is a plausible mechanism for producing striped jets. As we have shown, the resulting striped outflow possesses $\Phi_{\mathrm{MAD}}\sim 40$ with a jet power of $\sim10^{50}\,{\rm erg\,s^{-1}}$, consistent with LGRB observations.

The trajectory of the contours in the spacetime diagram indicates how rapidly the magnetic flux is advected inward or transported outward. \citet{2024MNRAS.532.1522J} provided analytic fits to the advection ($t_{\mathrm{adv}}$) and ejection (outward transport, $t_{\mathrm{ejc}}$) timescales of the poloidal loops
\begin{equation}
\begin{aligned}
    t_{\rm ejc} &= \frac{2}{a_1}(r^{1/2} - r_0^{1/2}) + t_0, \\
    t_{\rm adv} &= \frac{2}{3a_2}(r_0^{3/2} - r^{3/2}) + t_0,
\end{aligned}
\end{equation}
where $a_1=7.8\times 10^{-4}$ and $a_2=4.1\times 10^{-2}$ are fitting constants to the radial‑velocity profiles, derived under the assumption of turbulent driven accretion of a steady-state accretion disk, $r$ is the radial position of the magnetic loops at time $t$, and $r_0$ and $t_0$ denote their initial position and generation time, respectively. We overlay these timescales on the magnetic‑flux spacetime diagram of model \texttt{rc\_36\_B\_s} and find that they match the advection and ejection of the blue patches, which drive the system into the MAD state, reasonably well. The characteristic timescale for local magnetic‑field generation\footnote{The timescale for the local magnetic field generation, $t_{\mathrm{gen}} = 10\,\Omega_k^{-1}$, is not the same as the timescale for magnetic flux reversal that appears in butterfly diagrams in shearing-box simulations. The latter is $\sim 10$ orbital periods \citep{2010ApJ...713...52D, 2012MNRAS.422.2685S, 2016MNRAS.457..857S}, and is therefore roughly a factor of $\sim 6 - 7$ longer than the magnetic field generation time, and a factor of $\sim 50 - 60$ longer than the growth time of the MRI turbulence $\sim \Omega_k^{-1}$ \citep{1991ApJ...376..214B}.}, $t_{\mathrm{gen}}=10\, \Omega _k^{-1}$, with $\Omega _k=1/(r^{3/2}+a)$ being the Keplerian frequency, also matches the onset of the blue patches in the diagram. This suggests that the magnetic-field generation and advection in model \texttt{rc\_36\_B\_s} are consistent with the MRI-driven dynamo model \citep{2024MNRAS.532.1522J}.


However, in contrast to their results, our models require a significantly longer time to reach the MAD state, i.e., $t \gtrsim 10^{5}\,GM/c^{3}$ instead of $t \gtrsim 10^{4}\,GM/c^{3}$. This discrepancy may be related to the initial toroidal-field strength, which seeds the non-axisymmetric MRI and/or additional (M)HD instabilities that are responsible for driving the dynamo. In their model, the initial toroidal field has $\beta_{\hat{\phi}}^{-1} = B_{\hat{\phi}}^{2}/(2P)\sim 0.2$, whereas in our models the initial field is much weaker, $\beta_{\hat{\phi}}^{-1} \lesssim 10^{-2}$, and it takes time for the toroidal field to grow before approaching equipartition with the gas pressure so that $\beta_{\hat{\phi}}^{-1} \sim 0.1$, thereby delaying the onset of efficient dynamo action and, consequently, the transition to the MAD state.

\subsection{Magnetic Field Evolution} \label{subsec:fieldgrow}
We have shown that the MAD state emerges as coherent poloidal magnetic field loops are advected from large radii, $r \gtrsim 100\,GM/c^{2}$, onto the BH. This motivates us to examine the magnetic-field evolution in the outer regions of the disk. Figure~\ref{fig:dynamo_beta} (a) illustrates the emergence of dynamically important toroidal magnetic fields through the spacetime evolution of the average toroidal plasma beta, $\langle \beta_{\hat{\phi}}^{-1} \rangle =\langle B_{\hat{\phi} }^2\rangle /(2\langle P\rangle)$. The average of a quantity $Q$ is given as
\begin{equation}
    \langle Q \rangle = \frac{\int \rho \sqrt{-g}Qd\theta d\phi}{\int \rho \sqrt{-g}d\theta d\phi}.
\end{equation}
The white region corresponds to freely falling gas that has not yet formed a disk (i.e., the thermal pressure is negligible). The initial disk size matches the circularization radius (red) and is only weakly magnetized. As the toroidal field within the disk becomes dynamically important, the MRI and/or additional (M)HD instabilities are well resolved, and enable efficient angular-momentum transport and viscous radial expansion of the accretion disk \citep{shakura_theory_1976}. This radial expansion is a crucial step toward the onset of the MAD state, as the disk must grow to $r \gtrsim 100\,GM/c^2$ with $\beta_{\hat{\phi}}^{-1}(r \gtrsim 100\,GM/c^2) \sim 0.1$ for magnetic loops of sufficient size \citep{2021MNRAS.508.1241C} to be generated.

Figure~\ref{fig:dynamo_beta} (a) provides insight into why model \texttt{rc\_36\_B\_s} occupies the optimal regime for early jet launching (shown in Figure~\ref{fig:timeseries}), lying between models \texttt{rc\_18\_B\_s} and \texttt{rc\_72\_B\_s} --- it is the model in which the disk reaches the strongest toroidal magnetization at $r \sim 100\,GM/c^2$ the earliest, which could be understood as follows: When the disk is too compact, as in model \texttt{rc\_18\_B\_s}, $\beta_{\hat{\phi}}^{-1}$ is larger at the onset, but the magnetic loop generated is also smaller due to smaller disk size. The disk has to viscously expand to $\sim 100\,GM/c^2$ at which point larger-scale poloidal loops could be produced (see Figure~\ref{fig:fluxfunction}). By contrast, when the disk is larger, as in model \texttt{rc\_72\_B\_s}, it can produce larger magnetic loops, but the larger disk radius also reduces $\beta_{\hat{\phi}}^{-1}$, so the seed toroidal field takes longer to amplify, and the generated poloidal loops also require more time to be advected toward BH. As such, model \texttt{rc\_36\_B\_s} strikes the optimal balance between these two effects --- the disk is large enough to generate large‑scale poloidal loops that can be quickly advected toward the BH, but not so large that the seed toroidal field remains dynamically important throughout, allowing the production of poloidal loops that carry sufficient magnetic flux. Finally, when the seed toroidal field is weaker, as in model \texttt{rc\_36\_B\_w}, it requires a longer timescale for the seed toroidal field to amplify sufficiently to attain dynamical importance.

The aforementioned trend can be seen more clearly in Figure~\ref{fig:dynamo_beta} (b), where we show the timeseries of the toroidal magnetic‑field pressure, gas pressure, and the toroidal $\beta$, all measured at $r=100\,GM/c^2$. At this radius, model \texttt{rc\_36\_B\_s} remains consistently more magnetized than the other simulations during the early evolution ($t \lesssim 10^{4}\,GM/c^3$). The increase in gas pressure from a minimum marks the onset of disk formation, while the non‑negligible gas pressure prior to this rise is caused by disturbances from standing shocks. The initial growth of the toroidal field is largely determined by magnetic flux freezing, and we observe a roughly power-law scaling in its radial profile\footnote{Secondary processes such as standing-shock-induced turbulence could also play a role. However, we find that the large‑scale average ($\sqrt{\langle B_{\hat{\phi}}\rangle^{2}}$) is only a factor $\sqrt{10} \sim 3.2$ smaller than $\sqrt{\langle B_{\hat{\phi}}^{2}\rangle}$, implying that secondary processes are likely unimportant.}. The toroidal field increases as the disk forms, rising from its initial value of $10^{12}\,\mathrm{G}$ to $\sim 10^{13}\,\mathrm{G}$. The initial growth of $\langle \beta_{\hat{\phi}} \rangle$ is largely driven by the amplification of the toroidal field. Thereafter, the toroidal field\footnote{Note that $10^{25}$\,dyne corresponds to $1.6\times10^{13}$\,G.} saturates at $\sim 10^{12} - 10^{13}\,\mathrm{G}$ and then steadily decreases along with the gas pressure (due to viscous expansion), causing $\langle \beta_{\hat{\phi}} \rangle$ to remain at $\mathcal{O}(10^{-1})$.

While the general trends described above are likely robust across different models, we note that the mechanism leading to a persistent MAD state is inherently stochastic. Competing magnetic loops of opposite polarity introduce variability, making the establishment of a coherent large-scale field in the inner disk a random process. As a result, simulations with slightly different initial conditions may exhibit different jet onset times. It remains unclear what determines the persistence of the MAD state in our models. Nonetheless, in all cases considered here, the jet onset time remains robustly on the order of one to a few seconds.

\section{Discussion and Conclusions} \label{sec:conclude}

The production of Poynting-flux-dominated jets in collapsars requires strong poloidal magnetic fields threading the BH. Such fields may be inherited from the PNS at the time of collapse \citep{Gottlieb_2024}, although the efficiency of this process and the amount of flux ultimately retained by the BH remain uncertain. Meanwhile, differential rotation in massive stars predominantly generates toroidal magnetic fields. Therefore, in the absence of efficient flux inheritance, a dynamo is required to convert toroidal fields into the large-scale poloidal fields needed to power the jet. In this letter, we present the first three-dimensional GRMHD study, with initial conditions that closely follow those of pre-collapse stellar models computed by \texttt{MESA}, demonstrating that a dynamo can generate coherent, large-scale poloidal fields of sufficient strength to thread the BH and launch steady jets with a comparable power to what is inferred from LGRBs.

\subsection{Comparison with Previous Work} \label{subsec:compare}

\citet{2025PhRvD.111l3017S} presented the only prior numerical study of collapsar BZ jet formation via a dynamo. However, their simulations are axisymmetric, and the mean-field dynamo is imposed through a parameterized source term in the evolution equations for the electromagnetic fields \citep{2013MNRAS.428...71B,2014MNRAS.440L..41B,2021PhRvD.103d3022S,2021PhRvD.104f3026S}. As such, the dynamo is not captured self-consistently.

Our study differs in several key respects\footnote{See their Figure 3, 8 and Table 6.}: (i) the onset of jet launching in their simulations occurs at $\mathcal{O}(10)\,\mathrm{s}$ after disk formation, whereas we find jet launching within the first few seconds; (ii) although their horizon-threading flux can reach $\sim 10^{28}\,\mathrm{G\,cm^2}$, it frequently changes sign, with the typical timescale for the flux to remain at the same sign being less than a second. In contrast, our models reach saturated MAD states while maintaining a coherent magnetic field threading the BH horizon, unless a magnetic-flux inversion occurs; (iii) Whereas their Poynting luminosity rarely reaches $10^{50}\,\mathrm{erg\,s^{-1}}$ and fluctuates over several orders of magnitude, our jetted models have jet powers of at least $10^{50}\,\mathrm{erg\,s^{-1}}$ and remain steady unless a magnetic-flux inversion occurs; (iv) they assume a rapidly spinning BH ($a\geq +0.7$), while we use a moderately spinning BH, as required to produce an LGRB power with $\Phi_{\rm MAD}\gtrsim 10$ \citep{2023ApJ...952L..32G}. These comparisons may suggest that the dynamo we observe is more efficient at generating strong, large-scale, and coherent poloidal fields, and at launching more powerful jets. However, we do not rule out the possibility that these disagreements arise from systematic differences in the progenitor models. Nonetheless, our results underscore the importance of fully three-dimensional, self-consistent dynamo modeling for accurately capturing jet formation in collapsars.

\subsection{Implications for LGRBs} \label{subsec:implication}

This work, together with that of \citet{Gottlieb_2024}, addresses one of the longstanding unresolved problems in collapsar physics: the origin of the magnetic field that powers the jet. The dynamo operating in the collapsar-disk model may offer a robust alternative to the fossil PNS magnetic-field scenario. The fossil-field model relies on a specific ordering of multiple timescales, and even if the PNS possesses sufficiently strong magnetic flux, it remains unclear how much of that flux is retained by the BH. In contrast, as we showed in this study, the dynamo operating in the collapsar-disk model is more robust to variations in the initial rotation rate and magnetic-flux distribution. We emphasize that collapsar-disk dynamo and fossil-field scenarios are not mutually exclusive; even if the fossil-field mechanism operates, the dynamo within the collapsar disk will still proceed.

This result has two important implications for LGRBs. First, nearly any progenitor that forms an accretion disk may be capable of launching a jet, suggesting that rotation alone could be a sufficient condition for LGRB production. Second, the jet-launching time may depend on both the seed magnetic field strength and the circularization radius of the accreting plasma. Stronger magnetic fields and moderate circularization radii of $\sim 100\,\mathrm{km}$ appear to favor earlier jet launching. Nevertheless, our results suggest that jet launching typically occurs within the first few seconds after collapse.

\subsection{What Drives the Dynamo?} \label{subsec:whichdynamo}

The fact that an accretion disk forms (Figure~\ref{fig:density_jet}) and that Figure~\ref{fig:fluxfunction} shows consistency with \citet{2024MNRAS.532.1522J} suggests that the dynamo in the collapsar disk is driven by MRI. However, as we show in Appendix~\ref{sec:supp}, the spacetime diagram of the toroidal magnetic field is highly irregular, exhibiting only the weak periodic signatures that one would expect to exist in an MRI-driven dynamo of an accretion disk \citep{2019MNRAS.482..848D, 2020MNRAS.494.4854D}.

Such a seeming inconsistency could be due to the following reasons. First, in Appendix~\ref{sec:supp} we show that the collapsar disk is geometrically thick ($H/r \sim 0.3 - 0.5$ within $10 - 500\,GM/c^2$) due to the absence of neutrino cooling. It has been shown that geometrically thick disks produce more chaotic butterfly patterns \citep{2018ApJ...861...24H}. We note that our disks are thicker than those studied in \cite{2018ApJ...861...24H} and \cite{2024MNRAS.532.1522J}; it is therefore not surprising that our spacetime diagrams appear even more chaotic. Such thick disks ($H/r \sim 0.3 - 0.5$) are not very different from unstratified accretion disks, and unstratified shearing-box simulations are known to exhibit no clear butterfly pattern \citep{2008A&A...488..451L}, even though the MRI-driven dynamo is still in operation. Second, because the disk-jet system is highly misaligned and wobbling, and because magnetic loops can be generated off the midplane, the butterfly pattern will couple to the disk oscillations, further enhancing the apparent absence of ordered oscillations in the magnetic field. We also stress that the absence of a clear butterfly pattern does not necessarily imply the absence of dynamo action.

Finally, it is also possible that the dynamo observed in our collapsar model is not \textit{solely} driven by the MRI, but may also include contributions from other sources such as turbulence \citep{2012MNRAS.423.3148S, 2014MNRAS.445.3169O, 2024ApJ...962..158H, 2025ApJ...990..223B} induced by the standing shocks \citep{2010ApJ...713.1219E, 2012ApJ...751...26E, 2014MNRAS.445.3169O}, and/or large-scale disk wobbling which is known to couple negatively with the MRI \citep{fairbairn_interplay_2025}. However, the accretion flow, as shown in Appendix~\ref{sec:supp}, is highly azimuthally sheared, with an angular-velocity profile close to the Keplerian one. As such, there is no reason to doubt that shear plays an important role in the field generation mechanism, as would be expected from dynamo action within an accretion disk. Nonetheless, our consistency with the results of \cite{2024MNRAS.532.1522J} emphasizes that the MRI plays an important role in driving the dynamo in the simulations. Future work will better constrain whether the effects enumerated above are as important as, or more than, the MRI in driving the dynamo.

\subsection{Uncertainties} \label{subsec:uncertain}

There are still several unanswered questions arising from our simulations. One is how the initial toroidal field is amplified to the point of dynamical importance. We argued above that the early growth is largely governed by toroidal-flux-freezing \citep[see also][]{2005ApJ...631..446S}. It may seem surprising that model \texttt{rc\_36\_B\_w}, which begins with a weaker magnetic field, attains a toroidal-field strength and plasma-beta evolution comparable to the stronger-field models. However, we find that the weak-field model possesses only a factor of two less total toroidal flux compared to the strong-field model, further supporting the conclusion that the initial amplification is set primarily by flux-freezing, i.e., the collapsing gas is compressing a similar amount of toroidal flux down to $\sim 100\,GM/c^{2}$.

At first glance, this might suggest that the initial total toroidal flux available at large radii alone determines whether the collapsar-disk dynamo can generate large-scale poloidal loops capable of launching jets. In this picture, progenitors with insufficient toroidal flux (i.e., those near the lower bound of Figure~\ref{fig:mesa_initial_combine} (d) or \texttt{rc\_36\_B\_vw}) could fail to produce an LGRB. However, It remains uncertain whether the ability of the dynamo to generate large-scale poloidal loops depends primarily on the initial $\beta_{\hat{\phi}}$ or on numerical resolution. Although \citet{2024ApJ...960...97R} showed in their weak-field (b200) model that the dynamo's efficacy is largely independent of resolution, their high-resolution weak-field run only achieved a toroidal quality factor of $\sim 50$. In contrast, \citet{2020MNRAS.494.3656L} demonstrated that the dynamo's ability to produce large-scale poloidal fields disappears when the resolution is reduced. To verify this, one should perform higher-resolution simulations of the weak-field case (e.g., model \texttt{rc\_36\_B\_vw}).

Moreover, differential rotation inherited from the angular-momentum profile can twist the poloidal magnetic field (which we neglect here) into a dynamically important toroidal component during stellar core collapse \citep{2004ApJ...616.1086T, 2004ApJ...608..907Y, 2005ApJ...631..446S, 2006A&A...450.1107O, 2006MNRAS.370..501M}. However, an uncertainty arises because the progenitor's poloidal field is shown to be small-scale \citep{Gottlieb_2024}. Thus, whether the poloidal field is coherent on scales comparable to the MRI wavelength and/or length scales of relevant (M)HD instabilities, and therefore capable of generating toroidal fields of similar scales, remains uncertain. If the poloidal field loops in the progenitor star are sufficiently strong, they may prevent differential rotation from winding up and amplifying the toroidal field. In that case, the initial magnetic-field topology is better described as mixed toroidal-poloidal or purely poloidal. But note that a poloidal field strong enough to inhibit efficient TSD must also transport angular momentum outward, which would likely suppress disk formation. Additionally, the description of angular momentum transport in the progenitor star is not well understood, and several alternatives \citep{2019ApJ...881L...1F, 2019MNRAS.485.3661F, 2025ApJ...988..195S} exist that could impact the resulting magnetic field.

Two additional points are worth noting: (i) radiative cooling changes the generation of poloidal magnetic fields required for jet launching via a dynamo \citep{YourKey2025}. In collapsars, neutrinos are the primary cooling channel, which we neglect here, leaving it uncertain whether successful jet launching would occur in a more realistic collapsar environment, and (ii) model \texttt{rc\_36\_B\_w} exhibits only a brief MAD state, which is subsequently destroyed by the inward advection of a strong magnetic‑field loop of opposite polarity. Whether such behavior is imprinted in the LGRB light curve remains an open question. We will address these questions in future work.

\begin{acknowledgments}
We express our gratitude to Lorenzo Sironi for providing very helpful comments that improved the manuscript. We thank Mitch Begelman and Jim Fuller for providing helpful comments on our manuscript. We thank Prasun Dhang for useful discussions regarding dynamo processes. We also thank Masaru Shibata for useful discussions regarding the dynamo in collapsars. Ho-Sang (Leon) Chan acknowledges support from the Croucher Scholarship for Doctoral Studies by the Croucher Foundation. Ho-Sang (Leon) Chan also acknowledges support from the Flatiron Institute Center for Computational Astrophysics Predoctoral Program 2025 - 2026. This research used resources of the Argonne Leadership Computing Facility, a U.S. Department of Energy (DOE) Office of Science user facility at Argonne National Laboratory and is based on research supported by the U.S. DOE Office of Science-Advanced Scientific Computing Research Program, under Contract No. DE-AC02-06CH11357 (NeutronStarRemnants project).
\end{acknowledgments}

\bibliography{sample701}{}
\bibliographystyle{aasjournalv7}

\appendix
\begin{figure*}
    \centering
    \gridline{
    \fig{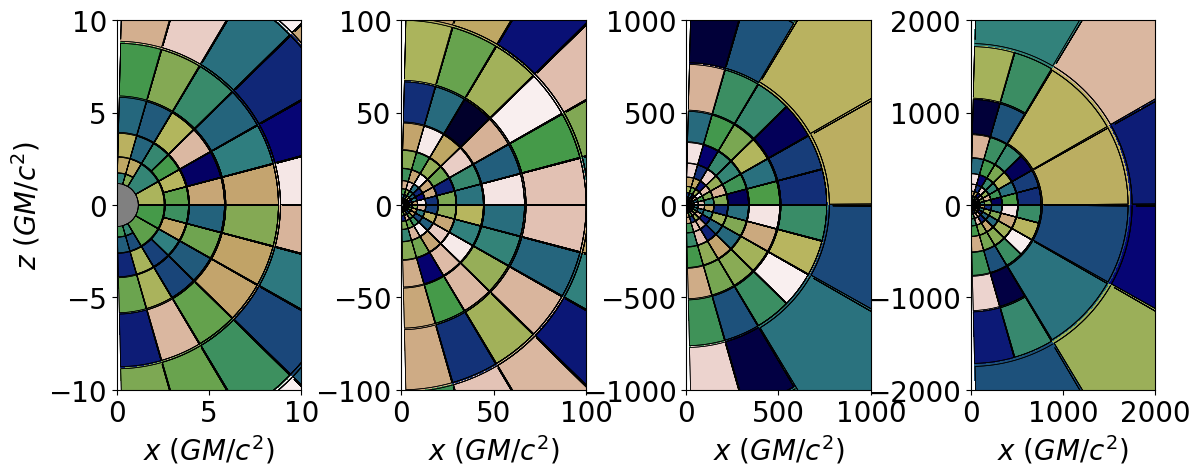}{1.0\textwidth}{(a)}}
    \gridline{
    \fig{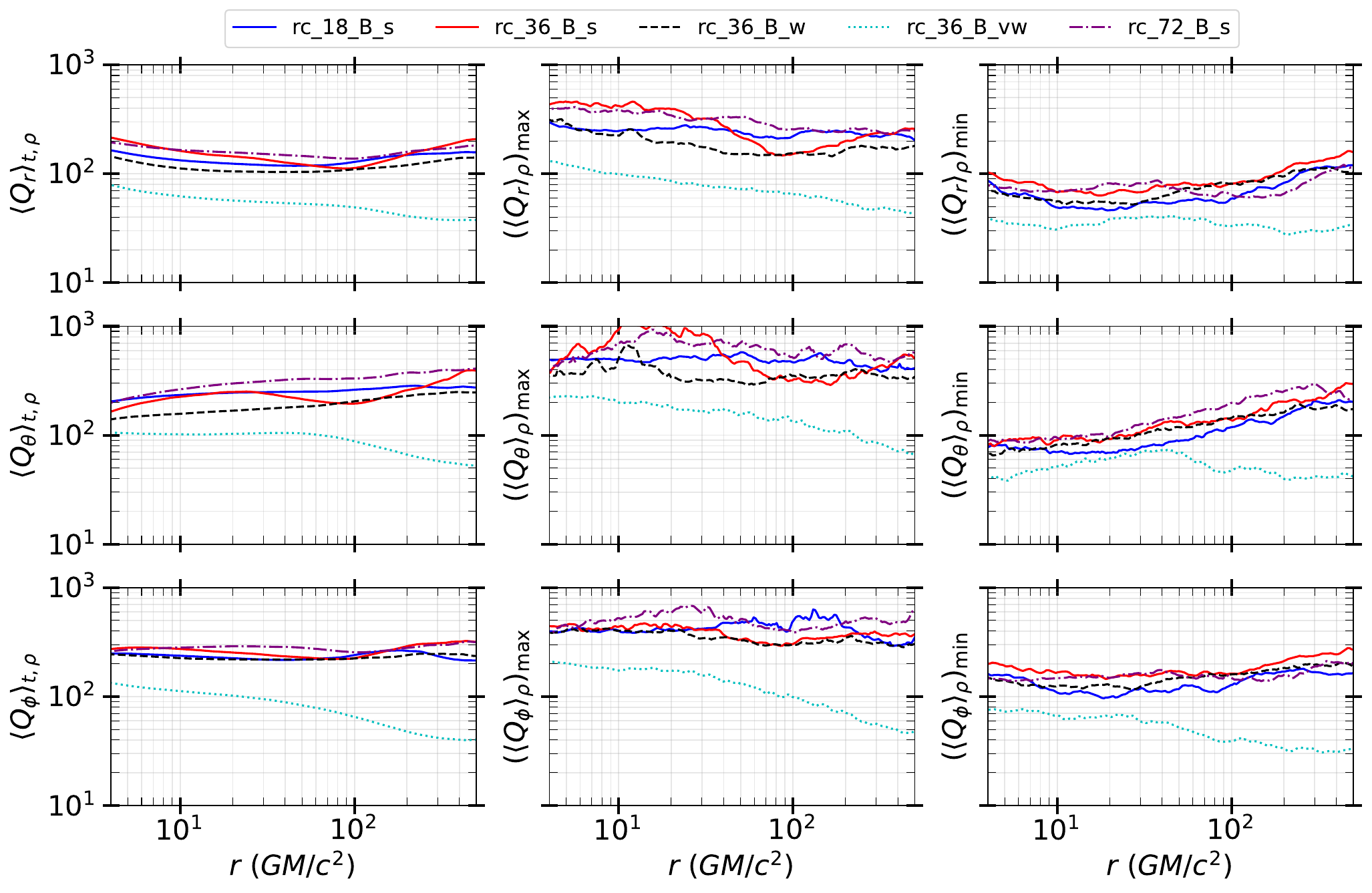}{1.0\textwidth}{(b)}}
    \caption{Supplementary Figures. (a) Mesh-block distributions for a range of plotting extents. The grey circle marks the event horizon. (b) MRI quality factors for all models within the SMR region $4 \leq r/(GM/c^2) \leq 500$. The first row shows the quality factor associated with the radial magnetic-field component: the left panel shows the time- and mass-averaged profiles, the middle panel shows the maximum values at each radius, and the right panel shows the minimum values at each radius. The second and third rows follow the same format but correspond to the $\theta$- and $\phi$-components of the magnetic field, respectively. All statistics use data with $t>1.5$\,$\mathrm{s}$ to ensure that the collapsar disk has formed self-consistently. \label{fig:supp}}
\end{figure*}
\begin{figure}[htb!]
    \centering
    \includegraphics[width=1.0\linewidth]{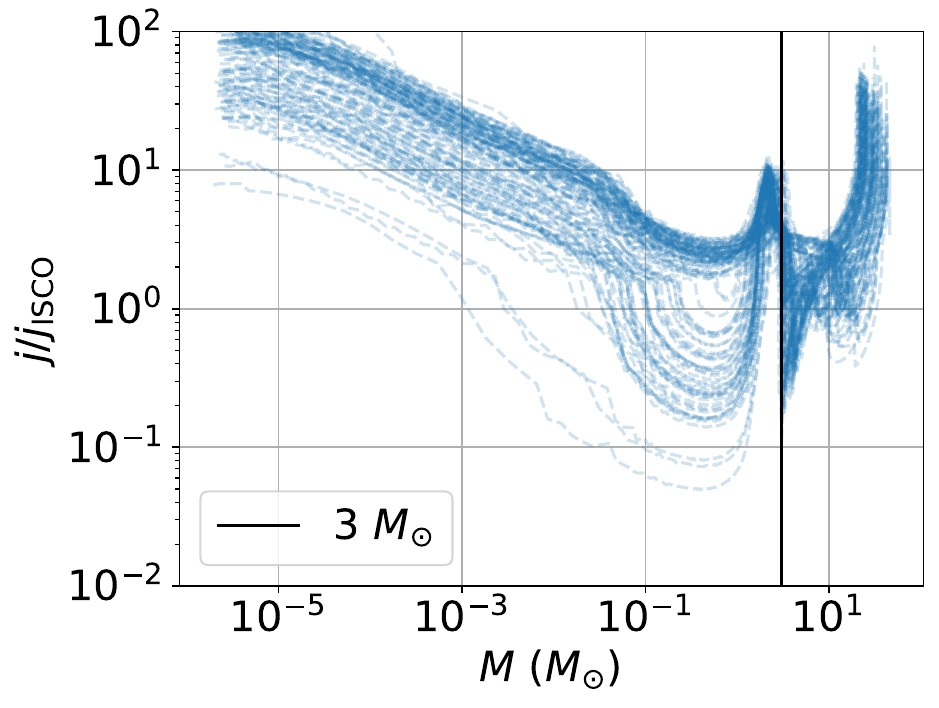}
    \caption{Ratio of specific angular momentum to the ISCO value as a function of mass coordinate. The ISCO quantity is computed self-consistently from the enclosed mass and its total angular momentum at each mass coordinate. Each curve corresponds to one stellar model. The vertical black line marks the location of the $3\,M_\odot$ mass coordinate. \label{fig:mj}}
\end{figure}
\begin{figure}[htb!]
    \centering
    \gridline{
    \fig{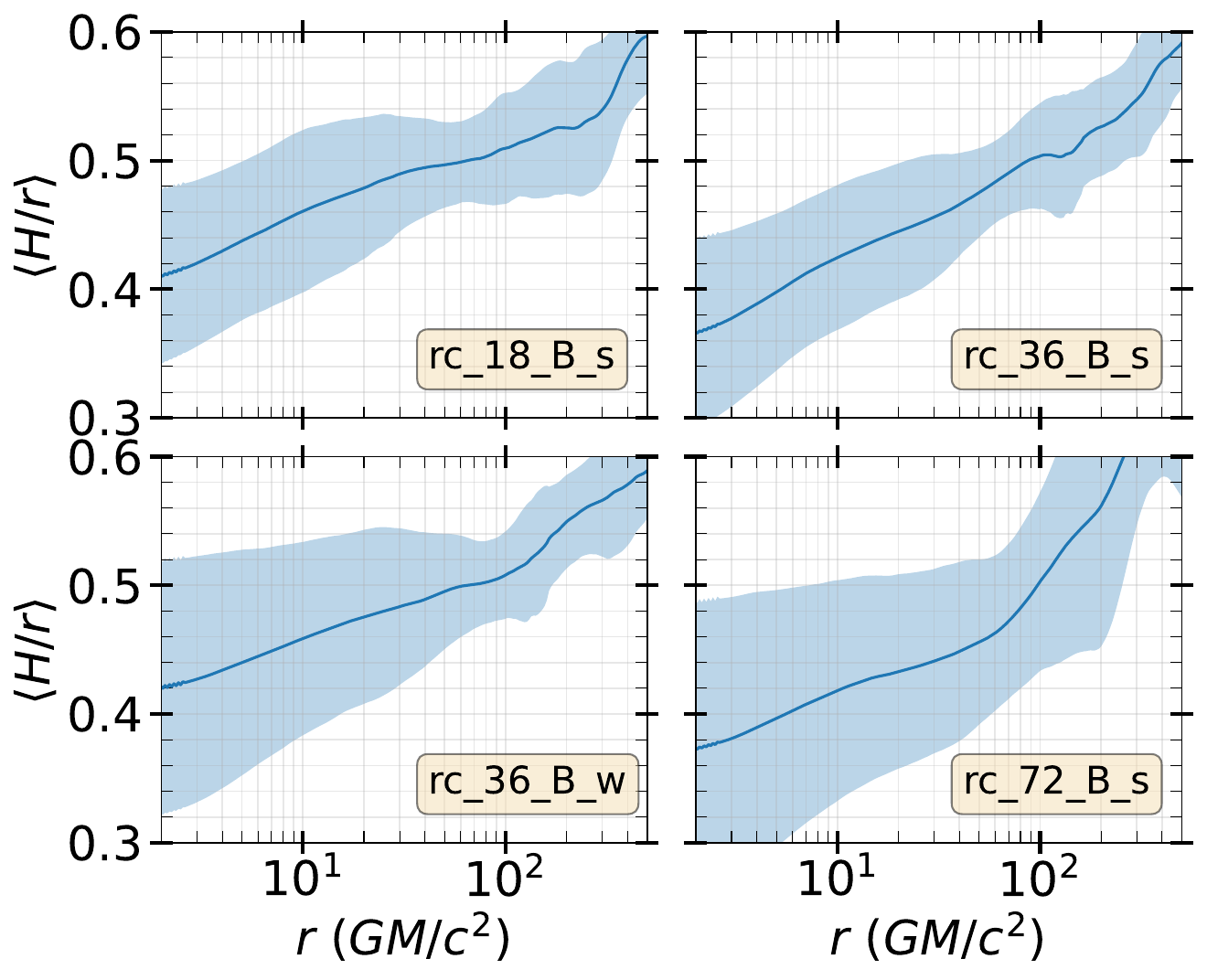}{1.0\linewidth}{(a)}}
    \gridline{
    \fig{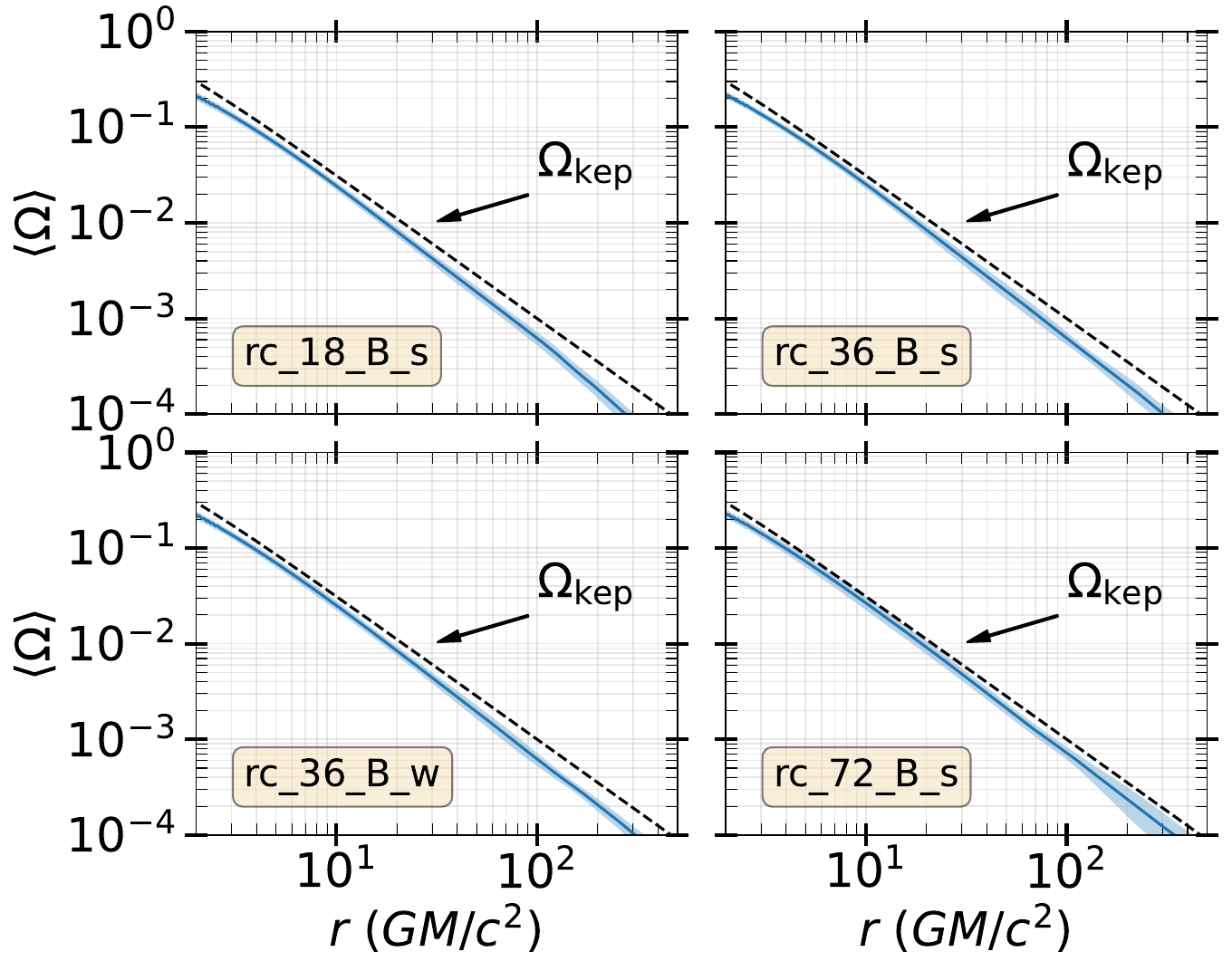}{1.0\linewidth}{(b)}}
    \caption{Panel (a): time-averaged (solid blue line) profile of the collapsar-disk scale height measured with respect to the midplane, where the shaded region denotes one standard deviation from the mean. Panel (b): same as panel (a), but for the mass-weighted angular-velocity profile. The black dashed line shows the Keplerian angular velocity, $\Omega_{\rm kep} = 1/(r^{3/2} + a)$. In both panels, the textbox indicates the model. The time-averaging is performed before the first jet is launched. \label{fig:average_profile}}
\end{figure}
\begin{figure*}
    \centering
    \gridline{
    \fig{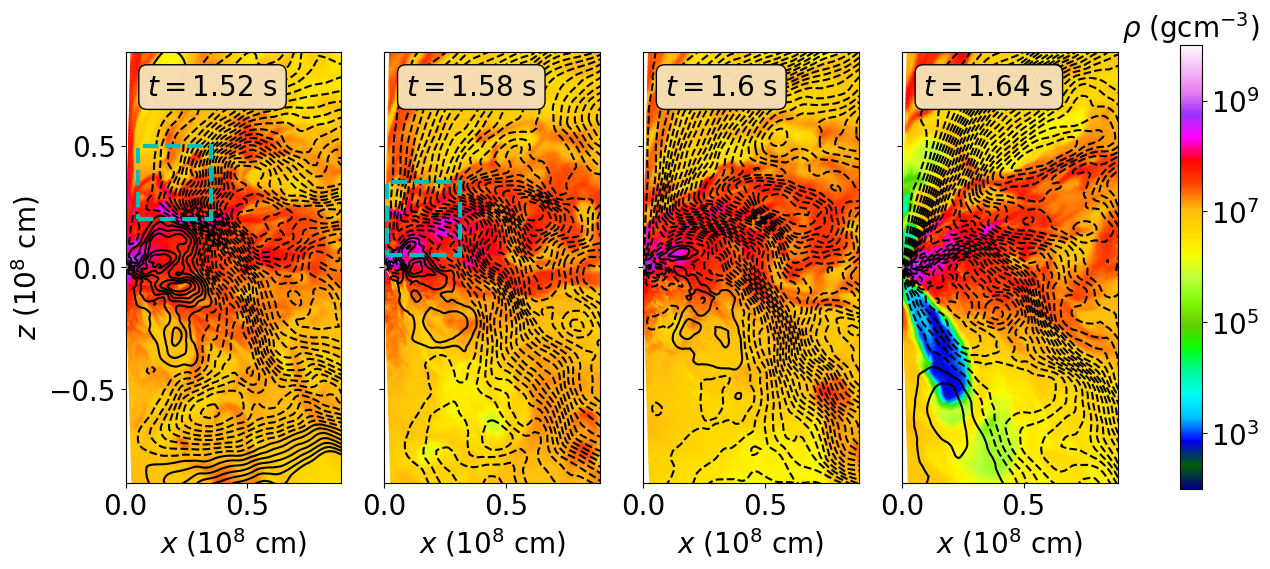}{1.0\textwidth}{(a)}}
    \gridline{
    \fig{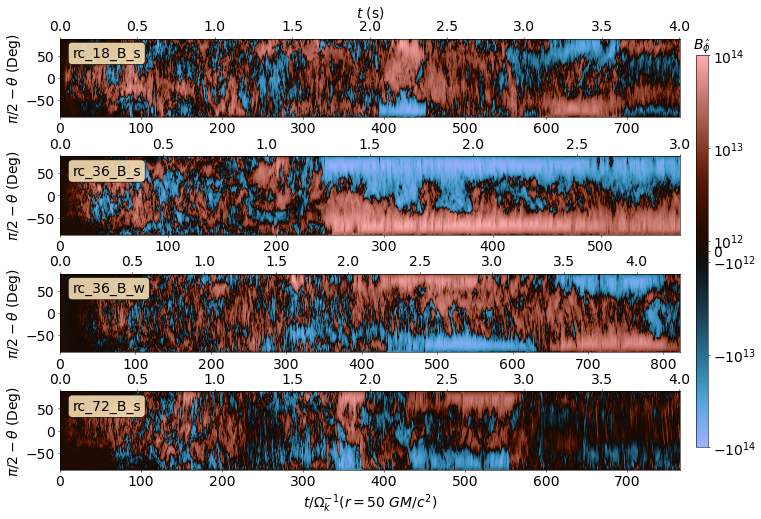}{1.0\textwidth}{(b)}}
    \caption{(a) Density snapshot in the $x-z$ plane at $\Phi =0^{\circ }$, with contours of the magnetic‑flux function (Equation~\ref{eqn:flux_func}, used here as a proxy for the averaged magnetic‑field lines) overlaid. We show model \texttt{rc\_36\_B\_w} at the instant when magnetic‑flux inversion occurs (see Figure~\ref{fig:fluxfunction}). Solid and dashed lines denote magnetic‑field loops of opposite polarity. Magnetic loops of polarity opposite to that threading the BH are highlighted at $1.52\,\mathrm{s}$ and $1.58\,\mathrm{s}$ with a rectangle. (b) Spacetime diagram of the azimuthally averaged toroidal magnetic field ($B_{\hat{\phi}}$) for all models except \texttt{rc\_36\_B\_vw}. The textbox in each row indicates the corresponding model. The magnetic field is measured at $50\,GM/c^2$. The time axis is shown in seconds and expressed in units of $\Omega_k^{-1}$ evaluated at $r = 50\,GM/c^2$. \label{fig:supp2}}
\end{figure*}

\section{GRMHD Simulations} \label{sec:grmhd}

To model the dynamo action and the resulting jet launching in collapsars, we solve the ideal GRMHD equations, where the equations (with speed of light $c = 1$, gravitational constant $G = 1$, and the BH mass $M= 1$) are
\begin{equation} \label{eqn:grmhd}
\begin{aligned}
    \partial_{t}(\sqrt{-g}\rho u^{t}) &= -\partial_{i}(\sqrt{-g}\rho u^{i}), \\
    \partial_{t}(\sqrt{-g}T^{t}_{\;\;\nu}) &= -\partial_{i}(\sqrt{-g} T^{i}_{\;\;\nu}) + \sqrt{-g}T^{\kappa}_{\;\;\lambda}\Gamma^{\lambda}_{\;\;\nu\kappa}, \\
    \partial_{t}(\sqrt{-g}B^{i}) &= -\partial_{j}[\sqrt{-g}(b^{j}u^{i} - b^{i}u^{j})], \\
    \partial_{i}(\sqrt{-g}B^{i}) &= 0.
\end{aligned}
\end{equation}
Here, $g_{\mu\nu}$ is the metric tensor, $\sqrt{-g}$ is the determinant of the metric, $\rho$ is the total gas density, $\Gamma$ is the Christoffel symbol, $B^{i}$ is the $3-$magnetic field, $u^{\mu}$ is the $4-$velocity, $b^{\nu}$ is the magnetic $4-$vector, and the stress-energy tensor is expressed as
\begin{equation}
    T^{\mu\nu} = (\rho h + b^{2})u^{\mu}u^{\nu} + (P + \frac{b^{2}}{2})g^{\mu\nu} - b^{\mu}b^{\nu},
\end{equation}
where $h = 1 + \gamma P/[(\gamma - 1)\rho]$ denotes the specific enthalpy, in which $P = u(\gamma - 1)$ is the pressure, $\gamma$ is the adiabatic index, and $u$ is the internal energy. We solve the spatial components of the GRMHD equations in the spherical-polar form of Kerr-Schild coordinates $(t, r, \theta, \phi)$ and advance the system in time using a second-order Runge-Kutta method with a Courant-Friedrichs-Lewy number of $0.9$. We impose floors on density $\rho$ and internal energy $u$ (in the code unit) such that $\rho > 10^{-10} r^{-2}$ and $u > 10^{-18} r^{-2 \gamma}$. In addition, we require that $b^{2}/\rho < 1000$, $b^{2}/u < 10000$, $u/\rho < 1000$, and the Lorentz factor measured in the orthonormal frame $\Gamma_{\rm norm} < 100$.

\section{Semi-analytic Evolution of the Free-falling Density Profile} \label{sec:semi-analytic}
To evolve the \texttt{MESA} profiles semi-analytically, we compute the free-fall time of the innermost edge (located at $r_{\rm FE}$) of the gas outside the BH mass coordinate
\begin{equation}
    t_{\rm fall} = \frac{1}{\sqrt{2GM}}\int_{r_{\rm FE}}^{r_{\rm BH}}\frac{\sqrt{r_{\rm BH}}\sqrt{r}}{\sqrt{r - r_{\rm BH}}}dr,
\end{equation}
in which we set $r_{\mathrm{BH}}=2$\,$GM/c^2$, the Schwarzschild radius. Given $t_{\mathrm{fall}}$ and the initial position $r$ of a fluid element, we then solve for the new position $r'$ of the free-falling fluid element
\begin{equation} \label{eqn:root}
    \sqrt{2GM}t_{\rm fall} = r^{3/2}\frac{\pi}{2} - r^{3/2}\sin^{-1}(\sqrt{\frac{r'}{r}}) + \sqrt{rr'(r - r')},
\end{equation}
where the expression on the right-hand side is obtained by integration. We solve this equation using a root-finding method. We also neglect the initial kinetic energy of the fluid element, as it is negligible compared to its gravitational energy.

The density profile can then be obtained by mass conservation
\begin{equation}
    \rho' = \rho \left( \frac{r}{r'} \right)^{2}\frac{dr}{dr'},
\end{equation}
where $dr/dr'$ is obtained by implicitly differentiating Equation \ref{eqn:root}.

\section{Resolution Study} \label{sec:resolution}
Figure~\ref{fig:supp} shows the mesh-block distribution for several choices of radial extent. Within $4 \leq r/(GM/c^2) \leq 500$, the refinement level reaches its maximum, ensuring uniform resolution across this region. Figure~\ref{fig:supp} also presents the MRI quality factors within the same radial range. We compute mass-averaged radial profiles of the MRI quality factor for the $r$, $\theta$, and $\phi$ components of the magnetic field. Because all cells in $4 \leq r/(GM/c^2) \leq 500$ are refined to the highest level, these statistics are robust. For each component, we show the time-averaged, time-maximum, and time-minimum profiles. All temporal statistics use data with $t > 1.5$\,s, after a collapsar disk has formed self-consistently.

Except for model \texttt{rc\_36\_B\_vw}, the average quality-factor profiles are strikingly similar across models, suggesting no systematic differences in MRI resolution associated with varying circularization radii. The average quality factors exceed $\sim 100$, indicating that the MRI is well resolved. The maximum quality factors ($\theta$ component) can exceed $1000$ in model \texttt{rc\_36\_B\_s}, which enters a persistent MAD state. In contrast, model \texttt{rc\_36\_B\_vw} exhibits lower quality factors, implying that MRI may be under-resolved in this model and potentially contributing to its failure to generate large-scale poloidal flux and launch a jet.

The minimum quality factors for the radial component occasionally fall to $\sim 40–80$, indicating brief intervals of marginal resolution. However, even the model that becomes MAD, \texttt{rc\_36\_B\_s}, reaches minimum values of $\sim 60$, and the minimum-quality-factor profiles show no systematic differences across models (again, except for model \texttt{rc\_36\_B\_vw}). These results suggest that the inability of some models to maintain a persistent MAD state is unlikely to stem from transient under-resolution of the MRI. Finally, the $\theta-$ and $\phi-$ component quality factors exceed those of the radial component, indicating that future improvements should focus on increasing the number of radial cells to further enhance MRI resolution.

\section{Supplementary Material} \label{sec:supp}
Figure~\ref{fig:mj} shows the ratio of the specific angular momentum to the ISCO angular momentum, computed using the enclosed mass and total angular momentum, as a function of mass coordinate for all progenitor models. The vertical line marks the $3\,M_\odot$ coordinate. Because the gas angular momentum exceeds the ISCO value at this location, a $3\,M_\odot$ BH cannot form promptly; instead, an accretion disk must form first. Once the proto-neutron star accretes beyond $3\,M_\odot$, it collapses into a BH that is already accompanied by a centrifugally supported disk. Thus, the seemingly prompt accretion-disk formation in our simulation is fully consistent with the angular-momentum structure of the progenitor model.

Figure~\ref{fig:average_profile} (a) and Figure~\ref{fig:average_profile} (b) show the time-averaged accretion-disk scale height (measured at the midplane) and the time-averaged angular-velocity profile, respectively. The absence of neutrino cooling renders the accretion disk geometrically thick, exhibiting $H/r \sim 0.3 - 0.5$. The large standard deviations (the shaded region bounding the solid line) reflect the highly wobbling, tilting disk behavior. Nonetheless, all models show a highly azimuthally sheared angular-velocity profile that is close to the Keplerian value.

Figure~\ref{fig:supp2} (a) shows an example of an interval in which magnetic‑flux inversion occurs, for which Figure~\ref{fig:fluxfunction} does not display any clear inward advection of magnetic flux. Our discussion focuses on model \texttt{rc\_36\_B\_w}, with the relevant instant highlighted in Figure~\ref{fig:fluxfunction}. At $t=1.52$\,s, the BH is still threaded by field lines of the original polarity (solid contours), but magnetic loops of opposite polarity (dashed contour, highlighted with a rectangle) in the upper hemisphere are ready to be transported toward the BH. By $t=1.58 - 1.60$\,s, that loop is dragged inward along the upper polar region toward the BH. By $t=1.64$\,s, sufficient opposite‑polarity magnetic flux has accumulated on the horizon to extract energy from the BH and launch a jet.

Figure~\ref{fig:supp2} (b) shows the spacetime diagram of the azimuthally averaged toroidal magnetic field for all models except \texttt{rc\_36\_B\_vw}. No clear periodic butterfly pattern is present, which we have discussed in the main text. We observe continuous sign reversals in the toroidal field with comparable magnitudes, indicating that toroidal-field generation driven by strong shear is active. For model \texttt{rc\_36\_B\_s}, a dominant sign of $B_{\hat{\phi}}$ emerges at $\sim 1.25$\,s and fills both the upper and lower polar regions. This coincides with the appearance of a poloidal magnetic loop that is subsequently advected onto the black hole, driving the system into the MAD state. After this transition, $B_{\hat{\phi}}$ maintains the same sign along the polar region. For model \texttt{rc\_36\_B\_w}, we observe similar behavior at $\sim 2$\,s, but at $\sim 3.5$\,s the toroidal field undergoes a sign reversal, corresponding to a magnetic-flux polarity reversal.




\end{document}